\documentclass[12pt]{iopart}
\usepackage{graphicx,bm,iopams,epsfig}

\def\brho{\mbox{\protect\boldmath $\rho$}}

\begin{document}

%% \date{February 26, 2002}

\def\reff#1{(\ref{#1})}
\newcommand{\be}{\begin{equation}}
\newcommand{\ee}{\end{equation}}
\newcommand{\<}{\langle}
\renewcommand{\>}{\rangle}

%%%  \ltapprox and \gtapprox produce > and < signs with twiddle underneath
\def\spose#1{\hbox to 0pt{#1\hss}}
\def\ltapprox{\mathrel{\spose{\lower 3pt\hbox{$\mathchar"218$}}
 \raise 2.0pt\hbox{$\mathchar"13C$}}}
\def\gtapprox{\mathrel{\spose{\lower 3pt\hbox{$\mathchar"218$}}
 \raise 2.0pt\hbox{$\mathchar"13E$}}}

\def\bsigma{\mbox{\protect\boldmath $\sigma$}}
\def\bpi{\mbox{\protect\boldmath $\pi$}}
\def\smfrac#1#2{{\textstyle\frac{#1}{#2}}}
\def\smhalf{ {\smfrac{1}{2}} }

\newcommand{\re}{\mathop{\rm Re}\nolimits}
\newcommand{\im}{\mathop{\rm Im}\nolimits}
\newcommand{\trace}{\mathop{\rm tr}\nolimits}
\newcommand{\fr}{\frac}

\def\Z{{\mathbb Z}}
\def\R{{\mathbb R}}
\def\C{{\mathbb C}}

\title{Coarse-graining strategies in polymer solutions}

\author{Giuseppe D'Adamo,$^1$ Andrea Pelissetto$^{2}$ and Carlo Pierleoni$^3$}

\address{$^1$ Dipartimento di Fisica, Universit\`a dell'Aquila, V. Vetoio 10,
   Loc. Coppito, I-67100 L'Aquila, Italy}
\address{$^2$ Dipartimento di Fisica, Sapienza Universit\`a di Roma and 
INFN, Sezione di Roma I, P.le Aldo Moro 2, I-00185 Roma, Italy}
\address{$^3$ Dipartimento di Fisica, Universit\`a dell'Aquila and 
CNISM, UdR dell'Aquila, V. Vetoio 10, Loc. Coppito, I-67100  L'Aquila, Italy}

\ead{giuseppe.dadamo@aquila.infn.it}
\ead{andrea.pelissetto@roma1.infn.it}
\ead{carlo.pierleoni@aquila.infn.it}

\begin{abstract}
We review a coarse-graining strategy (multiblob approach) 
for polymer solutions in which
groups of monomers are mapped onto a single atom (a blob) and 
effective blob-blob interactions are obtained by 
requiring the coarse-grained model to reproduce
some coarse-grained features of the zero-density isolated-chain structure.
By tuning the level of coarse 
graining, i.e.~the number of monomers to be mapped onto a single blob,
the model should be adequate 
to explore the semidilute regime above the collapse transition,
since in this case the monomer density is very small if chains 
are long enough. 
The  implementation of these ideas has been previously
based on a transferability hypothesis, which was not completely
tested against full-monomer results 
(Pierleoni {\em et al.}, {\em J.~Chem.~Phys.}, 2007, {\bf 127}, 171102). 
We study different models 
proposed in the past and we compare their predictions to full-monomer 
results for the chain structure and the thermodynamics in the range of 
polymer volume fractions $\Phi$ 
between 0 and 8. We find that the transferability assumption has 
a limited predictive power if a thermodynamically consistent 
model is required. We introduce 
a new tetramer model parametrized in such 
a way to reproduce not only zero-density intramolecular and intermolecular 
two-body probabilities, but also some intramolecular 
three-body and four-body distributions. We 
find that such a model correctly predicts three-chain 
effects, the structure and the thermodynamics up to 
$\Phi \simeq 2$, a range considerably larger than that obtained with 
previous simpler models using zero-density potentials. 
Our results show the correctness of the 
ideas behind the multiblob approach but also that more work is needed to 
understand how to develop models with more effective monomers 
which would allow us to explore the semidilute regime at larger
chain volume fractions. 
\end{abstract}

\maketitle
\section{Introduction}
Polymer solutions show a wide variety of behaviors, depending on chain length,
density, and temperature.\cite{deGennes,Doi,desCloizeauxJannink,Schaefer}
In the dilute regime the isolated chain radius of gyration $R_g$ is the 
relevant 
length scale and the properties of the solution can be described in terms of 
single-chain properties and of the solvent-quality parameter. 
The radius of gyration scales as  $R_g=bL^{\nu}$, where
$b$ is the monomer characteristic length (Kuhn segment),
which depends on chemical details and temperature, $L$ is 
the number of monomers per chain, and $\nu$ is a universal exponent.
In the good-solvent regime\cite{Clisby:2010p2249} $\nu = 0.587597(7)$, 
while $\nu = 1/2$ (with logarithmic corrections 
\cite{desCloizeauxJannink,Schaefer}) for $\theta$-solvents.
The semidilute regime is entered when chains start to overlap, but still
the monomer density is small. If $c = N/V$ is 
the polymer concentration --- $N$ is the number of chains and 
$V$ the volume of the system under consideration --- and 
$c_m = c L$ is the monomer concentration, the semidilute regime
is characterized by $c > c^*$ (or equivalently by $\Phi > 1$, where 
$\Phi = c/c^*$ is the polymer volume fraction) and $c_m \ll 1$, where 
$c^*=3/(4\pi R_g^3)$ is the overlap concentration. Note that 
$c_m=(3/4\pi b^3) \Phi L^{1-3\nu}$. Hence, when increasing $\Phi$, 
increasingly longer polymer chains are needed to ensure the semidilute 
condition $c_m \ll 1$. If $c_m$ is not small, one enters the 
concentrated or melt regime.
%% In the semidilute regime chains overlap and the
%% isolated-chain statistics (either self-avoiding or ideal) is retained
%% only at length scales smaller 
%% than the de Gennes-Pincus correlation length $\xi$, which scales as 
%% $\xi=R_g \Phi^{-\gamma}$ with\cite{deGennes,Doi,desCloizeauxJannink,Schaefer} 
%% $\gamma=\nu/(3\nu-1)$. 
%% In good-solvent conditions $\gamma\simeq 0.770$ while at 
%% $\theta$ conditions $\gamma=1$. 
%% At length scales beyond $\xi$ the screening of the excluded-volume interactions 
%% leads to ideal (Gaussian) statistics in all cases. 
%% In the melt regime, $\xi$ is
%% comparable with the monomer size and 
%% the chain statistics is ideal at all length scales. 

From this discussion it appears that very long polymers are necessary
to obtain a genuine semidilute regime over several orders of 
magnitude in chain density. For this reason, 
simulations of semidilute solutions of linear chains, even at the level of  
generic lattice or bead-spring models with implicit solvent, are 
quite expensive and have been limited to not too long chains and 
not too high densities.
\cite{Muller:2000p2000,Cavallo:2006p784,Bolhuis:2001p268,%
Louis:2002Physica,Yan:2000p2257,Pelissetto:2005p296} Moreover, 
in many complex situations, polymers only constitute one species in the 
solution, making full-monomer simulations even more difficult. 
In these cases a modelling at length scales of the order of the 
polymer size is often sufficient to provide the relevant thermodynamic and 
structural informations on the solution.
For instance, to determine the phase behavior of polymer-colloid mixtures
in the colloid regime,
\cite{PoonWCK:2002p1993,Fuchs:2002p2258,Tuinier:2003p2259,Mutch:2006p2260}
a detailed microscopic model of the polymers is not necessary. It is 
enough to use coarse-grained models which retain the essential thermodynamic 
(long-wavelength) behavior of the polymer solution.\cite{Bolhuis:2002p267}
Other examples are block copolymers for which 
the self-assembling of the chains in supramolecular aggregates of 
various shapes and sizes is ubiquitous. The description of the physical 
behavior of the self-assembled phases only requires a modelling at the 
mesoscopic level rather than at the microscopic (monomer) level.
\cite{Pierleoni:2006p159,HansenJP:2006p2248,SG-07,GP-10} This is 
the realm of self-consistent field-theoretical methods which have proved to 
be very effective to describe the physics of concentrated solutions and 
melts of homopolymers and block-copolymer blends. 
\cite{Bates:1990p759,Cavallo:2006p784} 

Coarse-grained models for soft condensed matter systems have received 
much attention in the last two decades.\cite{Likos:2001p277} 
In the simplest approach one maps polymer chains onto point particles
interacting by means of the pairwise potential of mean force between the 
centers of mass of two isolated polymers.
\cite{Dijkstra:1999p2142,Likos:2001p277} 
This potential is of the order of $2k_BT$ at full 
overlap,\cite{Grosberg:1982p2265,Dautenhahn:1994p2250}
has a limited range of the order of $3 R_g$ and
is very well represented by a 
linear combination of few Gaussian functions. 
\cite{Bolhuis:2001p268,Pelissetto:2005p296}
Such a model is however limited to the dilute regime $\Phi\lesssim 1$, in which 
many-body interactions\cite{Bolhuis:2001p288} can be neglected. 
This limitation was overcome 
ten years ago in a seminal work,\cite{Louis:2000p269,Bolhuis:2001p268}
which eliminated the complexity of the many-body interactions by introducing
a density-dependent pair potential, which is unique according to Henderson
theorem \cite{Henderson:1974p2091} and
reduces to the potential of mean force in the limit of zero density.
%% As shown by Louis {\em et al.},\cite{Louis:2000p269} 
%% the density-dependent potential
%% can be determined quite accurately by using standard 
%% integral-equation methods of liquid-state theory. Since the interactions
%% are soft, the hypernetted chain (HNC) equation can be 
%% successfully used to invert the centre-of-mass radial distribution function 
%% at finite density, as obtained at the full-monomer level, to determine the 
%% density-dependent pair interaction between the coils. Surprisingly enough, the 
%% form of the potential depends only slightly, but in a non-trivial way, on 
%% density. This dependence is essential to reproduce the 
%% des Cloizeaux scaling behaviour of the equation of state (EOS) of the 
%% solution.\cite{deGennes,desCloizeauxJannink}

This work has paved the way to the use of soft effective particles to 
represent polymer coils in complex situations such as in modelling 
colloid-polymer mixtures.\cite{Bolhuis:2002p267}
However, density-dependent potentials are 
difficult to handle. Care is needed to derive the correct thermodynamics 
\cite{HansenMcDonald,Stillinger:2002p2125,Louis:2002p2193} and to compute 
free energies and phase diagrams.\cite{Likos:2001p277}
Also their use in non-homogeneous situations is 
cumbersome since the interaction should depend on the local density which is 
not known beforehands and some kind of self-consistent procedure should be 
developed. Furthermore, representing polymers as soft spherically symmetric 
particles is not always appropriate. For instance, in studying polymers 
adsorbed on surfaces, like polymer brushes or polymer-coated colloids, 
it is clear that the anchorage to the surface breaks the rotational symmetry 
of the chains, an effect that must be taken into account in any accurate 
coarse-grained model. A further example is in modelling solutions of 
A-B block copolymers which cannot be represented as soft particles interacting 
by a spherically symmetric pair potential.
\cite{Pierleoni:2006p159,HansenJP:2006p2248,SG-07,GP-10}

In principle those limitations can be overcome by switching to a model at a 
lower level of coarse graining, i.e. by mapping a long linear polymer to a 
short linear chain of soft effective monomers (called ``blobs'' in the 
following). Such model retains some internal degrees of freedom, which 
allow more flexibility in chain geometry as necessary, for instance,
in anisotropic systems. 
Moreover, in the semidilute regime, this model is expected to 
allow the use of density-independent blob-blob interactions, 
since the local density of the blobs can always be kept small by increasing the 
number of blobs per chain.  Indeed, if chains of $L$ monomers are 
partitioned in $n$ effective blobs of $m = L/n$ monomers each, the local
concentration of the blobs is $c_b = c n$. The blob overlap
concentration is given by $c^*_b = 3/(4 \pi r_g^3)$, where 
$r_g$ is the radius of gyration of the blob. If we assume that 
$r_g = b m^\nu$, where $b$ is the Kuhn length that appears in the 
scaling of the radius of gyration (this relation, though not
exact, is a very good approximation), we obtain 
\begin{equation}
{c_b\over c_b^*} = {3\over 4\pi b^3} {n c\over m^{3\nu}} = \Phi n^{1-3\nu}
\approx \Phi n^{-0.763}.
\end{equation}
Hence, for any polymer volume fraction $\Phi$, since $\nu > 1/3$ above the 
collapsed phase,  one can choose $n$ so that $c_b/c_b^* < 1$, i.e., so that 
blobs do not overlap. In this regime, the size of each blob is 
approximately density independent, hence each blob can be replaced by
an effective single atom of fixed size in a coarse-grained representation. 
Moreover, one expects the many-body
interactions among blobs of different chains to be negligible. Hence, the 
parametrization of the intermolecular 
interactions among the chains in terms of 
zero-density pair potentials should 
be reasonably accurate. Conversely, one can use the $n$ blob model 
with zero-density intermolecular pair potentials up 
to $\Phi\lesssim n^{3\nu - 1}$. For larger concentrations many-chain 
intermolecular interactions come into play.

The main problem in this 
approach is how to obtain the intramolecular interactions, i.e., the 
potentials among the blobs of the same chain. The problem is trivial for a 
dumbbell model ($n=2$), where the blob-blob interaction is just the 
potential of mean 
force, but becomes increasingly difficult when increasing the number $n$ of 
blobs per chain. Indeed, the intramolecular effective interaction, which
is simply minus the logarithm of the joint distribution of the blob positions, 
is inherently a many-body interaction, which cannot be represented 
as a hierarchy of two-body, three-body, etc. terms, at variance with
what happens for the intermolecular potentials. In that case the 
small-density expansion gives the pair potential at lowest order, the 
three-body potential at next-to-leading order, and so on. 
Therefore, approximations must be introduced in trying to reproduce some 
features of the underlying full-monomer system.

As a first attempt in this direction, Pierleoni {\em et al.}
\cite{Pierleoni:2007p193} introduced a multiblob model, 
referred to as model M1 in the following, for homopolymers 
in good-solvent conditions. They started from the 
intramolecular and intermolecular potentials appropriate for dumbbells 
(a two-blob molecule), a
problem that can be solved as explained in 
Addison {\em et al.}\cite{Addison:2005p225} 
The potentials for chains with more blobs were then obtained from the 
dumbbell potentials using a transferability hypothesis.
This model has the correct scaling behavior for 
good-solvent polymers in dilute and semidilute solutions,
\cite{Pierleoni:2007p193} including the excluded-volume screening 
at large length scales with density. However, its prediction for the EOS of the 
solution is incorrect,\cite{Pelissetto:2009p287} 
shading serious doubts on the correctness of the 
transferability assumption for the potentials. 
A modified model, referred to as model M2 in the following, was also 
introduced.\cite{Pelissetto:2009p287} In this model, for each number of blobs,
the parameters of the potential 
are tuned to match full-monomer results for the EOS.
Although successful in reproducing the thermodynamics, 
model M2 is inherently different from M1 in that full-monomer results 
at finite density are needed to tune the model parameters, which is an evident 
limitation of this approach in more complex situations.

In this paper, beside comparing in detail results for models M1 and M2 to
full-monomer predictions in a wide range of polymer concentrations in the 
semidilute regime, we introduce a new coarse-grained model for semidilute 
solutions. In this model we map a single chain on a tetramer, i.e., a chain of 
four blobs, in such a way to reproduce all two-body, and some three-body and 
four-body distributions of an isolated good-solvent polymer at the full 
monomer level. In order to accomplish this program we use bonding, bending and 
torsional angle potentials, plus additional 1-3 and 1-4 central potentials 
which we obtain with the Iterative Boltzmann Inversion (IBI) procedure.
\cite{Schommers:1983p2118,MullerPlathe:2002p2127,Reith:2003p2128}
Furthermore, a single intermolecular pair 
potential between blobs of different tetramers is obtained in such a way 
to reproduce the center-of-mass radial distribution function between two 
isolated chains at the full-monomer level. This model provides 
correct results for the EOS up to reduced densities 
$\Phi\simeq 2$, a considerably larger range of densities with respect 
to simpler models. The tetramer model presented here is a first successful 
attempt to realize the multiblob program: 
building a multiblob coarse-grained model 
based on zero-density potentials which is able to provide the correct 
single-chain structure and EOS at finite density.

Our approach is very close  to the coarse-graining procedure 
applied by Fritz {\em et al.}\cite{Fritz:2009p1721} to polystyrene. 
Similar methods 
were also used for coarse-grained simulations of typical benchmark
chains like polycarbonates and polystyrene in a melt, 
\cite{Milano:2005p2251,Carbone:2008p2254} although in these works 
the potentials were 
fixed by requiring the coarse-grained model to reproduce structural 
properties at fixed pressure and temperature --- hence potentials
also depend on thermodynamic variables, as is the case for the 
density-dependent potentials.\cite{Louis:2000p269,Bolhuis:2001p268}
In some coarse-graining procedures
also thermodynamic information was taken into account to fix the 
potentials, see, e.g., Rossi {\em et al.}\cite{Rossi:2010p2256} 
and references therein.
We should also mention the approach of Vettorel {\em et al.}
\cite{Vettorel:2010p1733}, which extends previous work on a 
single-blob model.\cite{Murat:1998p1980}
In this multiblob model each blob carries internal degrees of freedom,
to account for the density profile of the underlying full-monomer 
subchains. However, the interactions are not derived {\em ab initio}
in the coarse-graining procedure, but are obtained by using 
phenomenological arguments. Finally, we mention the work of 
Clark {\em et al.} \cite{CG-10} which applies integral-equation methods to a 
coarse-grained model appropriate for polymer melts.

The paper is organized as follows. In section \ref{sec.2} 
we present 
the general formalism behind any coarse-graining procedure and we report 
our specific methodology to derive the tetramer model. In 
section \ref{sec.3} we compare the structure and the thermodynamic 
behavior predicted by the tetramer model with those 
(referred to as full-monomer results in the following) obtained 
by using lattice polymer models with a large number of 
monomers ($L\gtrsim 1000$) --- hence appropriate to obtain the 
universal, scaling behavior --- both at zero and at 
finite density in the semidilute regime. 
In section \ref{sec.4} 
we report results for the coarse-grained models M1 and M2 and 
compare them with the tetramer and the full-monomer data. Finally, 
we collect our conclusions and perspectives in the last section. 
In the appendix we give  universal predictions for 
the blob radius of gyration, an important quantity to obtain a meaningful 
comparison between any coarse-grained model and the underlying full-monomer 
model. 

\section{The blob model} \label{sec.2}

In order to obtain the coarse-grained blob model (CGBM), 
one works in the zero-density
limit and determines in successive steps the intramolecular potentials, 
the two-body intermolecular potentials, then, at least in principle, the 
three-body, four-body, etc. intermolecular potentials. In an exact mapping 
all $k$-body intermolecular interactions should be considered. 
However, as discussed in
the introduction, higher-order intermolecular interactions can be neglected 
if one only considers small densities $\Phi \lesssim \Phi_{\rm max}$, where 
$\Phi_{\rm max} \sim n^{3\nu-1}$ increases (for a given level of approximation)
with the number $n$ of blobs.

\subsection{The blob representation of the polymer}  \label{sec2.1}

In the multiblob approach, the basic object is the ``blob", 
which is a subchain of the polymer. Suppose we wish to partition a
polymeric chain of $L$ monomers into $n$ blobs of $m=L/n$ monomers each.
If the monomer positions are given by
$\{ {\bf r}_1,\ldots, {\bf r}_L\}$, one first defines the
blob positions ${\bf s}_1,\ldots, {\bf s}_n$ as the
centers of mass of the subchains of $m$ monomers, i.e.
\begin{equation}
 {\bf s}_i = {1\over m} \sum_{k=m(i-1)+1}^{mi}  {\bf r}_k.
\end{equation}
For the new coarse-grained chain $\{{\bf s}_1,\ldots, {\bf s}_n\}$
one defines several standard quantities. First, one defines its
radius of gyration
\begin{equation}
{R}_{g,b}^2 = {1\over 2 n^2} \sum_{i,j} ({\bf s}_i - {\bf s}_j)^2 .
\label{Rgb-def}
\end{equation}
Such a quantity is always smaller than ${R}_g$, since
\begin{equation}
{R}_g^2 = {R}_{g,b}^2 + {1\over n} \sum_i {r}_{g,i}^2,
\label{Rg-Rgb}
\end{equation}
where ${r}_{g,i}$ is the radius of gyration of $i$-th blob:
\begin{equation}
{r}_{g,i}^2 = {1\over 2 m^2}
   \sum_{k,l= m(i-1)+1}^{mi}  ({\bf r}_k - {\bf r}_l)^2.
\end{equation}
The ratios $R_{g,b}^2/R_g^2$ and $r_{g,i}^2/R_g^2$ of their averages 
\footnote{Note that here we use the same notation for the squared radius of 
gyration of a single-chain configuration 
and for its statistical average over all chain 
conformations. When we will need to distinguish between the two quantities,
the average squared radius of gyration will be indicated as 
$\langle R^2_g\rangle$.}
over the polymer configurations are universal, hence independent
of the nature of the underlying polymer model as long as $L$ is large enough. 
The average of the blob squared radius of gyration ${r}_{g}$ defined by
   $ {r}_{g}^2 = {(1/n)} \sum_i {r}_{g,i}^2$
scales quite simply with $n$ in the zero-density limit. 
As discussed in \ref{App-rgblob}, for all $n\ge 4$ we have 
quite precisely
\begin{equation}
{\hat{r}_{g}^2\over \hat{R}_g^2}  = 1.06 n^{-2\nu},
\label{scaling-rg2}
\end{equation}
an expression we will use extensively in the present work 
(here and in the following we will use a hat to indicate zero-density 
quantities).

Beside the radius of gyration, we can consider the bond-length distributions
(all blob distributions depend on the number $n$ of blobs, which is 
implicit in the notation)
\begin{equation}
  P_{ij}(r) = \langle \delta(|{\bf s}_i - {\bf s}_j| - r) \rangle,
\end{equation}
where $\langle\cdot \rangle$ is the statistical average over all 
chain conformations, which satisfy the normalization conditions
\begin{equation}
  \int_0^\infty dr\ P_{ij}(r) = 1.
\end{equation}
They depend on the chosen length scale. As it is standard
in renormalization-group analyses of polymer behavior, the relevant quantities
are the adimensional combinations $\hat{R}_g P_{ij}(r)$.
For $L\to \infty$ they
converge to universal, hence model-independent,
functions $f_{ij}(\rho)$ with $\rho = r/\hat{R}_g$, which are normalized as
\begin{equation}
  \int_0^\infty d\rho\ f_{ij}(\rho) = 1 .
\end{equation}
Note that, as usual, scaling functions depend only on the adimensional
combination $\rho = r/\hat{R}_g$.

In this paper we will also consider the adimensional
intramolecular distribution function 
\begin{equation}
g_{\rm intra}({r}) = 
   {2 \hat{R}^3_g \over n (n-1)} \sum_{i<j} 
   \langle \delta^{(3)}({\bf s}_i - {\bf s}_j - {\bf r})\rangle.
\label{gintra-def}
\end{equation}
For large $L$, $g_{\rm intra}({r})$ converges to a universal 
function $G_{\rm intra}(\rho)$, $\rho=r/\hat{R}_g$, which is related to the 
bond-distribution functions defined above by
\begin{equation}
G_{\rm intra}(\rho) = {1 \over 2 \pi n (n-1)} \sum_{i<j} 
   {f_{ij}(\rho)\over \rho^2}.
\label{Gintra-def}
\end{equation}
Note that the the ratio $R_{g,b}^2/\hat{R}_g^2$ 
is simply related to the second moment of $g_{\rm intra}(r)$ [in the 
scaling limit to that of $G_{\rm intra}(\rho)$]. 
For $L\to \infty$ we have
\begin{equation}
{R_{g,b}^2\over \hat{R}_g^2} = 
   {(n-1)\over 2 n} \int \rho^2 
    G_{\rm intra}(\rho) d^3\brho.
\label{Rg-Gintra}
\end{equation}
Beside two-site distributions, one can define three-site
correlation functions
\begin{equation}
P_{i,jk}({\bf r}_1,{\bf r}_2) =
    \langle \delta^{(3)}({\bf s}_i - {\bf s}_j - {\bf r}_1)
            \delta^{(3)}({\bf s}_i - {\bf s}_k - {\bf r}_2) \rangle
\end{equation}
--- the corresponding adimensional combinations
 $\hat{R}_g^6 P_{i,jk}({\bf r}_1,{\bf r}_2)$
converge to universal functions of ${\bf r}_1/\hat{R}_g$ and
${\bf r}_2/\hat{R}_g$ --- and, analogously, four-site, five-site, etc.
correlations.

As a check of the quality of our results we shall often consider the 
distribution of $R_{g,b}$. More precisely, for each polymer configuration
we consider the corresponding radius $R_{g,b}$ and the adimensional 
ratio $R_{g,b}/{\langle\hat{R}_g^2\rangle}^{1/2}$, 
where $\langle\hat{R}_g^2\rangle$ 
is the average of the squared radius of gyration over the polymer 
configurations. 
The corresponding distribution 
\begin{equation}
P_{R,b}(q_b) = \left\langle 
   \delta\left({R_{g,b}\over \sqrt{\langle\hat{R}_g^2\rangle}} - q_b\right)
    \right\rangle
\label{distPRb}
\end{equation}
is universal in the large-$L$ limit. Note that this distribution function
cannot be written in terms of the bond-length distributions, but is 
instead a particular $n$-blob correlation since $R_{g,b}$ depends on the 
positions of all blobs.

\subsection{The coarse-grained model} \label{sec2.2}

In the CGBM the basic object is a polyatomic molecule 
with $n$ atoms located in ${\bf t}_1,\ldots,{\bf t}_n$. 
All length scales are expressed in terms of $\hat{R}_g$, 
hence potentials and distribution functions depend on the 
adimensional combination $\brho = {\bf t}/\hat{R}_g$.
The intramolecular potentials are determined by requiring all 
{\em adimensional} distributions to
be identical in the polymer model and in the CGBM at zero density.
For instance, we require 
\begin{equation} 
\langle \delta(|{\bf t}_i - {\bf t}_j|/\hat{R}_g - \rho) \rangle_{CGBM} = 
    f_{ij}(\rho),
\end{equation}
where $\langle \cdot \rangle_{CGBM}$ is the average over all 
single-chain CGBM configurations and $f_{ij}(\rho)$ are the universal functions
defined above, which are computed in the polymer model. 

In principle, the determination of the intramolecular potential is
straightforward.
First, one determines the $n$-body blob distribution in the polymer 
model at zero density:
\begin{equation}
P_n({\bf r}_{12},\dots,{\bf r}_{1n})=
  \left\langle 
   \prod_{k=2}^{n}\delta^{(3)}
    \left({\bf s}_k - {\bf s}_1 - {\bf r}_{1k}\right)\right\rangle,
\end{equation}
where the average is over all single-polymer conformations. The 
adimensional combination $\hat{R}_g^{3(n-1)}P_n$ converges for $L\to \infty$
to a universal distribution:
\begin{equation}
\hat{R}_g^{3(n-1)} P_n({\bf r}_{12},\dots,{\bf r}_{1n}) = 
    f_n\left(\brho_{12}, \ldots,
             \brho_{1n} \right).
\end{equation}
where $\brho_{ij} = {\bf r}_{ij}/\hat{R}_g$.
The CGBM intermolecular potential is then
\begin{equation}
\beta V(\brho_1,\ldots, \brho_n) = 
   -\log f_n\left(\brho_2 - \brho_1, \ldots, \brho_n - \brho_1\right),
\label{pot-nbody}
\end{equation}
where $\brho_i = {\bf t}_i/\hat{R}_g$. By definition, this choice 
ensures that the distribution of the $n$ atoms in the CGBM is identical 
to the distribution of the $n$ blobs in the polymer model. Hence the 
intramolecular structure is exactly reproduced.  Note that potential
(\ref{pot-nbody}) is an intrinsically $n$-body interaction and thus there is 
no natural method to represent it as a sum of two-body, three-body, etc., 
terms. Because of the universality of the function $f_n$, the potential 
is independent of the polymer model and is valid for any polymeric 
system under good-solvent conditions.

The radius of gyration ${R}_{g,CGBM}$ of the CGBM molecule differs from the 
polymer radius of 
gyration ${R}_g$ but agrees instead with ${R}_{g,b}$.
It is important to take this difference into account when comparing 
finite-density results. For polymers, the behavior is universal once densities
are expressed in terms of the polymer volume fraction
\begin{equation}
   \Phi = {4\pi\over 3} \hat{R}_g^3 {N\over V},
\label{Phip-def}
\end{equation}
where $N$ is the number of polymers contained in the volume $V$. 
Full-monomer results should be compared with results obtained in the CGBM
at volume fractions
\begin{equation}
   \Phi_b = {4\pi\over 3} \hat{R}_g^3 {N_b\over V},
\label{Phib-def}
\end{equation}
where $N_b$ is the number of CGBM molecules. Note that $\hat{R}_g$ and not 
$\hat{R}_{g,b}$ appears in the definition of $ \Phi_b$. 
Since $\hat{R}_{g,b}/\hat{R}_g$ converges to 1 as $n$ increases,
for $n$ large, say $n\gtrsim 30$, this conceptual difference is not 
relevant in practice. 
In our case, instead, since we consider $n=4$, 
it is crucial to use the correct definition, that is the quantity $\Phi_b$.

Once the intramolecular potentials are determined, one must determine
the intermolecular potentials, which must be such to reproduce the 
potentials of mean force in the polymer model. Note that, in order to have 
an exact mapping of the polymer model onto the CGBM, not only should 
pair potentials be considered, but also three-body, four-body, etc. interactions
should be included.\cite{Bolhuis:2001p288,Pelissetto:2005p296}
However, as we already discussed in the introduction,
as $n$ increases, these many-body interaction
potentials are expected to become smaller, 
so that the CGBM with only pair potentials
should be accurate in a density interval which widens with increasing $n$.

\subsection{Determination of the four-blob 
CGBM intramolecular potentials} \label{sec.2.2}

In order to have an exact mapping of the polymeric system onto the 
$n$-blob CGBM, one should consider an $n$-body 
intramolecular potential, which, for $n>2$, 
can be expressed in terms of $3(n-2)$ scalar combinations of 
the positions of the blobs because of rotational and translational invariance.
The complexity increases rapidly with $n$ and for this reason we 
decided to consider the case $n=4$, which allows us to limit the number of 
approximations needed and, at the same time, allows us to 
go beyond the dilute regime up to $\Phi\approx 2$-3. However, 
even for $n$ as small as 4, an exact determination of the 
intramolecular potential requires considering a 
function of 6 independent variables, which is far too complex in practice.

Thus we have used a limited set of interactions. 
The intramolecular interactions have been modelled by introducing 
six different potentials, each of them depending on a single scalar variable.
This choice is arbitrary, but, as we will show in the following, it is 
particularly convenient and works quite well.
First, we consider  a set of bonding pair potentials: 
atoms $i$ and $j$ of the tetramer interact with a pair potential 
$V_{ij}(\rho)$ with $\rho = |{\bf t}_i - {\bf t}_j|/\hat{R}_g$. 
Because of symmetry we have 
$V_{13}(\rho) = V_{24}(\rho)$ and $V_{12}(\rho) = V_{34}(\rho)$, so that 
there are only four independent potentials to be determined. 
Then, we consider
a bending-angle potential $V_b(\cos \beta)$
and a torsion-angle potential $V_t(\theta)$, where $\beta$ and 
$\theta$ are defined as
\begin{eqnarray}
&& \cos\beta_i = {\Delta {\bf t}_i \cdot \Delta {\bf t}_{i+1} \over 
              |\Delta {\bf t}_i| |\Delta {\bf t}_{i+1}|},
\label{bending} \\
&& \cos\theta_i = 
   {(\Delta {\bf t}_i \times \Delta {\bf t}_{i+1}) \cdot 
    (\Delta {\bf t}_{i+1} \times \Delta {\bf t}_{i+2}) \over 
   |\Delta {\bf t}_i \times \Delta {\bf t}_{i+1}|
   |\Delta {\bf t}_{i+1} \times \Delta {\bf t}_{i+2}| },
\label{torsion}
\end{eqnarray}
with $\Delta {\bf t}_i = {\bf t}_{i+1} - {\bf t}_i$. 
Note that in the tetramer there are two bending angles, which are 
equivalent by symmetry, and a single torsion angle. 
This particular form of the potential set is inspired 
by the usual modelling of bonded interactions in macromolecules. 
However, in that context one only considers a bonding potential between
atoms which are first neighbors along the chain, a bending and a torsional 
term. Instead, our parametrization 
includes interactions between atoms that are not 
neighbors along the chain, thereby taking into account to some
extent cross-correlations
among  different degrees of freedom. We note that the bending potential
and the torsion potential involve three and four atoms, respectively,
and thus allow us to introduce some of the three-body and four-body interactions
present in the exact parametrization.

Since we are using a limited set of interactions, not all distributions of
the internal degrees of freedom can be exactly reproduced by the CGBM.
We must therefore choose the distributions which we wish to be identical 
in the polymer and in the CGBM case. 
Given our choice of interaction potentials, 
it is natural to use the adimensional bond-length distributions 
$\hat{R}_g P_{ij}(r)$ and the distributions of the bending and torsion 
angle [in the blob representation of the polymer model, 
these angles are defined by replacing
${\bf t}$ with ${\bf s}$ in Eqs.~(\ref{bending}) and (\ref{torsion})], 
which are particular three-body and four-body correlation functions. 
If we indicate collectively the six potentials to be determined with 
$V_{i}(x_i)$, the (adimensional) distributions of the $x_i$ variables with 
$P_i(x_i)$ in the CGBM and with $P_{i,FM}(x_i)$ in the full-monomer 
case, the potentials should be such that $P_{i}(x_i) = P_{i,FM}(x_i)$.
The universal (i.e., model-independent) distributions $P_{i,FM}(x_i)$ in 
the polymer case 
have been determined by performing simulations of self-avoiding walks on
a cubic lattice. To detect scaling corrections, we consider chains of 
length $L=2100$, 4100 (the corresponding blobs have $L/4 = 525, 1025$
monomers, respectively). The (adimensional) distributions obtained in the 
two cases agree within errors, indicating the absence of finite-length effects. 

We determine the potentials of the CGBM 
by using the Iterative Boltzmann Inversion (IBI) scheme. 
\cite{Schommers:1983p2118,MullerPlathe:2002p2127,Reith:2003p2128}
In this approach the effective
interactions which reproduce the 
target structural quantities are determined iteratively.
The potentials of mean force of the corresponding 
full-monomer distribution, $ -\ln P_{i,FM}(x_i)$,
have been chosen as initial guesses for all interactions except for 
$V_{14}(r)$. 
For $V_{14}(r)$ we have assumed a simple Gaussian potential:
we use a Gaussian approximation of the 
potential of mean force between two polymer centers of mass,
rescaling the width of the Gaussian
with the ratio of the radii of gyration
of the blob and of the entire chain.

\begin{figure}
\begin{center}
\begin{tabular}{c}
\epsfig{file=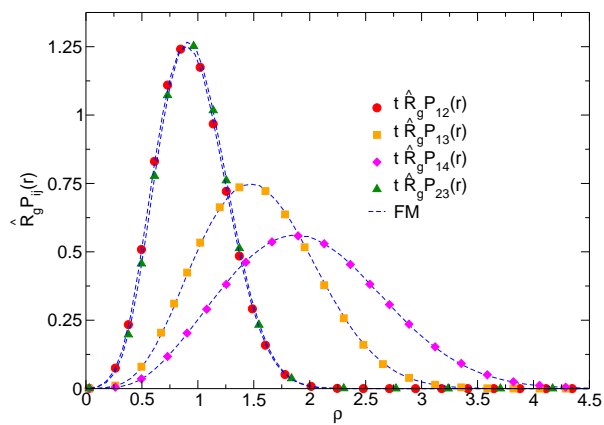,angle=-0,width=8truecm} 
    \hspace{0.5truecm} \\
\end{tabular}
\end{center}
\caption{Adimensional two-body distributions $\hat{R}_g P_{ij}(r)$ 
as a function of $\rho = r/\hat{R}_g$.
We report full-monomer results (dashed line) and the results for the tetramer
CGBM with the potentials obtained by means of the IBI procedure.
}
\label{dist-radial.fig}
\end{figure}

\begin{figure}
\begin{center}
\begin{tabular}{cc}
\epsfig{file=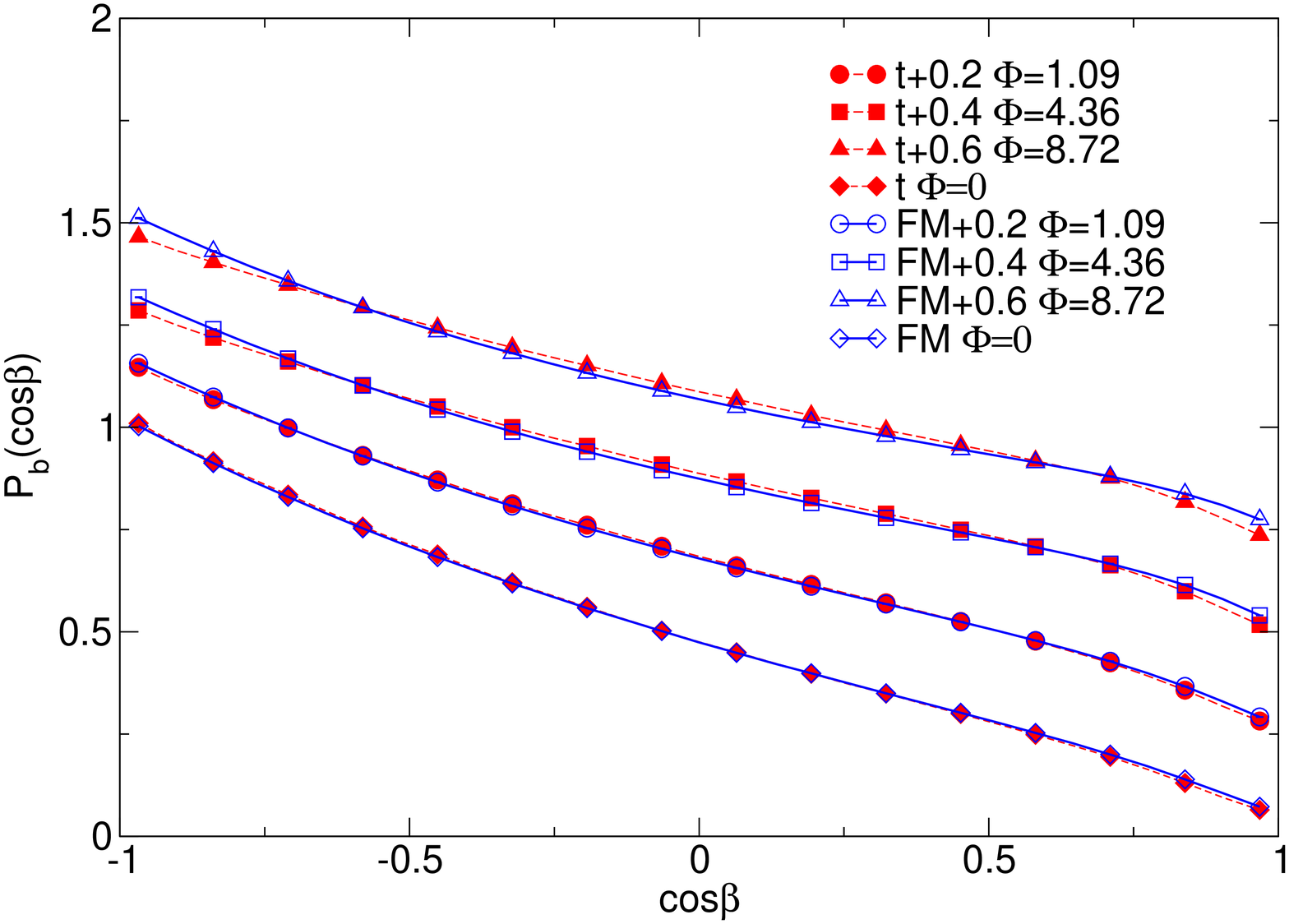,angle=-0,width=7truecm}  &
\hspace{0.5truecm}
\epsfig{file=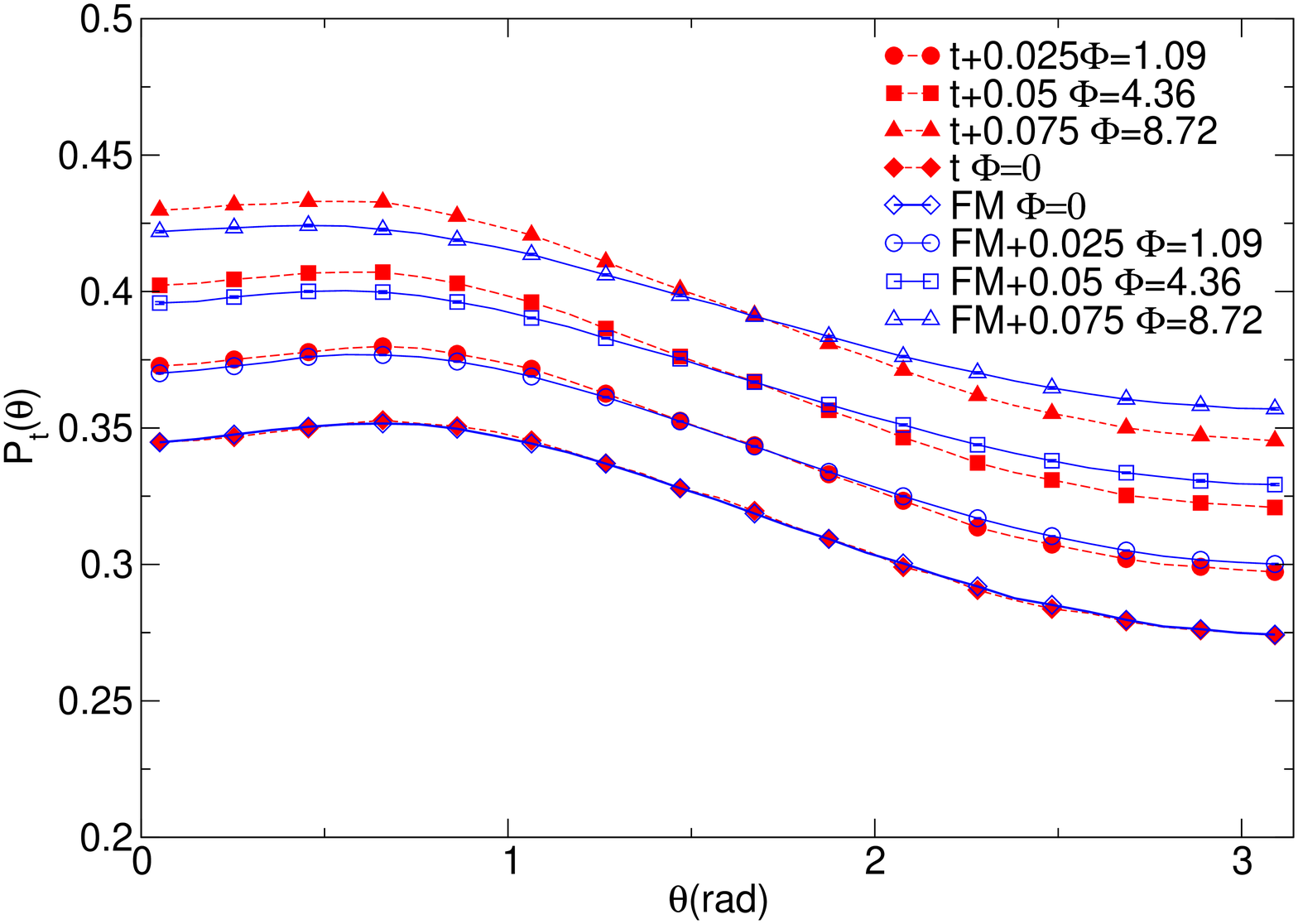,angle=-0,width=7truecm} \\
\end{tabular}
\end{center}
\caption{Distributions of the torsion angle $\theta$ (right) and 
of the cosine of the bending angle $\beta$ 
(left) as a function of $\theta$ (in radians) and of 
$\cos\beta$, respectively.
We report full-monomer results (FM, open symbols and solid line) and the results 
for the tetramer CGBM (t, closed symbols and dashed lines) with the 
potentials obtained by means of the IBI procedure 
at $\Phi = 0$, $\Phi = 1.09$, $\Phi = 4.36$ and $\Phi = 8.72$.
For sake of clarity, results at different densities are shifted upward by constant values reported in the legend. 
}
\label{angoli.fig}
\end{figure}

\begin{figure}
\begin{center}
\begin{tabular}{c}
\epsfig{file=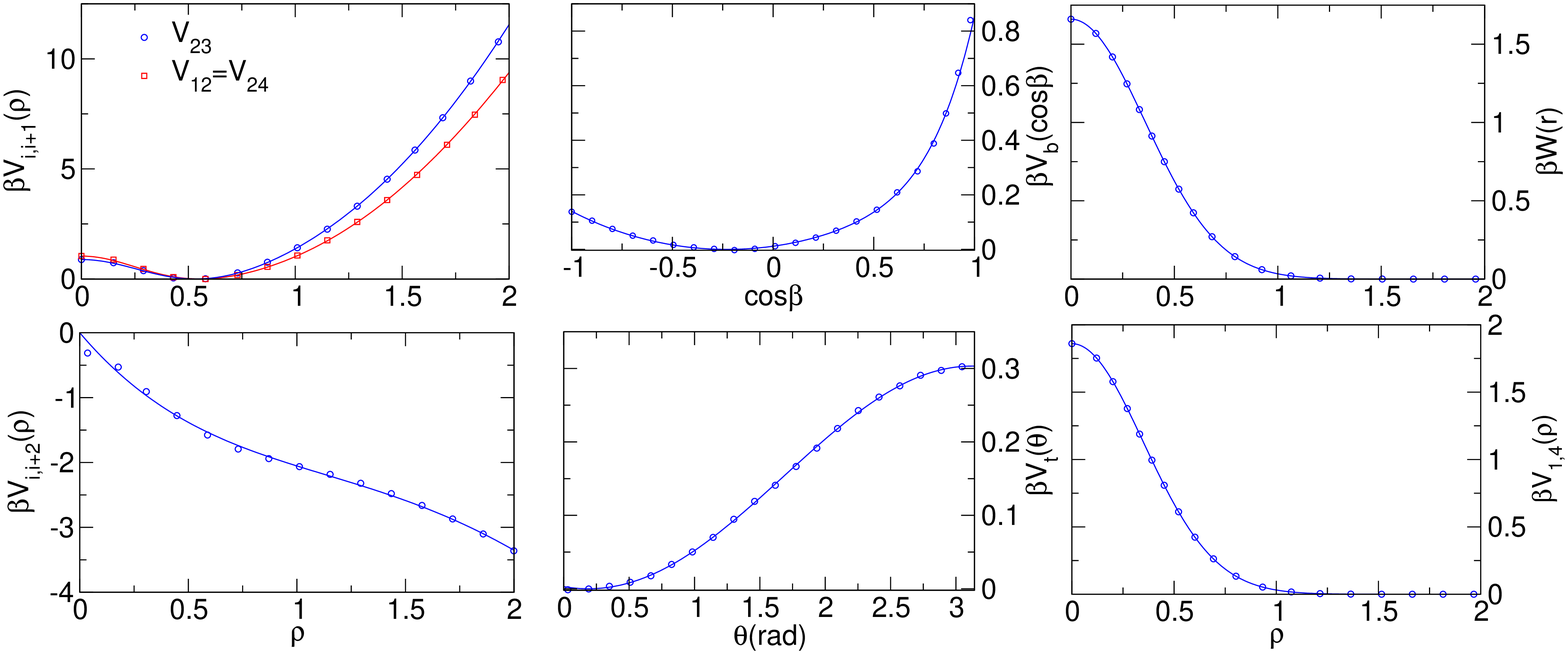,angle=-0,width=16truecm} 
    \hspace{0.5truecm} \\
\end{tabular}
\end{center}
%% \vspace{1cm}
\caption{Potentials for the tetramer CGBM. In the left column we show
the pair potentials between neighbors along the chain (top) and
between next-to-nearest atoms (bottom), 
as a function of $\rho = r/\hat{R}_g$.
In the central column we report the bending-angle potential as a function of 
$\cos \beta$ (top) and the torsion-angle potential as a function of
$\theta$ (in radians) (bottom). In the right column we plot
the intermolecular potential (top) and 
the potential between the first and last atom of the chain (bottom)
as a function of $\rho = r/\hat{R}_g$.
The symbols represent the numerical results obtained for a discrete set of 
values of $\rho$, the lines are interpolations.
}
\label{potenziali.fig}
\end{figure}

At the end of the optimization procedure, the bond and angle distributions
are reproduced quite precisely, see figures~\ref{dist-radial.fig} and 
\ref{angoli.fig}.
The potentials obtained are reported in figure~\ref{potenziali.fig}.
The pair potentials $V_{12}(\rho)$ and $V_{23}(\rho)$
are very similar, indicating that end effects are not very important. 
They have a minimum for $\rho\approx 0.5$ and are very soft at the 
origin: $V(0) - V(0.5) \approx (0.9$-1.0)$\ k_BT$. For $\rho\to\infty$
they increase quite rapidly and for $2\lesssim \rho \lesssim 3 $
they approximately behave as $\rho^{2.2}$ ($V_{12}$) and $\rho^{2.4}$ 
($V_{23}$). The potential between next-nearest neighbors is continuously
decreasing and hence it penalizes configurations in which the two atoms are 
close. Finally, $V_{14}(\rho)$ appears to be irrelevant for $\rho\gtrsim 1$.
As for the bending-angle potential, it penalizes configurations with 
$\beta < 90^\circ$, while it is essentially flat for $\beta > 90^\circ$:
the potential has a minimum for $\beta\approx 100^\circ$ and 
$V_b(180^\circ) - V_b(100^\circ) \approx 0.14 k_BT$. Finally, the torsion 
potential turns out to be quite flat: it increases with $\theta$ and changes 
only by $0.3 k_BT$, going from $0^\circ$ to $360^\circ$. 

The results for the potentials are quite interesting since they indicate the 
presence of an effective hierarchy among the different contributions.
The pair potentials between neighbors are the most important. For instance,
for typical configurations, say, for $0.4 \lesssim \rho_{12},\rho_{23}
\lesssim 1.5$, see figure~\ref{dist-radial.fig}, the potentials
$V_{12}(\rho_{12})$ and $V_{23}(\rho_{23})$ vary significantly, by 
4$k_BT$-5$k_BT$. Instead, for typical distances 
$0.7\lesssim \rho_{13} \lesssim 2.2$, 
the potential $V_{13}(\rho_{13})$ varies much less, approximately 
by 2$k_B T$, while in the typical interval $\rho_{14}\gtrsim 1$, 
$V_{14}(\rho_{14})$ varies only by 0.03$k_BT$. Clearly, the relevance of the 
interactions decreases as the chemical distance between the atoms 
increases, even though interactions between distant atoms can never 
be neglected, otherwise one would model a random-walk chain and not a 
polymer under good-solvent conditions. 

The bending-angle potential varies at most by $k_BT$ and is thus 
less relevant than the bonding potentials,
while the torsion potential only provides 
a small correction. This seems to indicate that the relevance of the 
interactions decreases with the number of atoms involved, so that,
when the number $n$ of atoms increases, one can safely neglect 
higher-body interactions.

It is important to stress that the pair potentials $V_{12}(\rho)$ 
and $V_{23}(\rho)$
are somewhat different from the potentials one would obtain 
by using the transferability hypothesis as suggested 
by Pierleoni {\em et al.}\cite{Pierleoni:2007p193} 
If we use Eq.~(\ref{scaling-rg2}) to relate $\hat{r}_g$ to $\hat{R}_{g}$, 
we would obtain 
a transferability potential (see the expression reported in 
the caption of figure 1 of Pierleoni {\em et al.}\cite{Pierleoni:2007p193})  
\begin{equation}
V_{tr}(\rho) = 1.92 \exp(-3.85 \rho^2) + 
            0.534 (2.19 \rho - 0.73)^2,
\end{equation}
where $\rho = r/\hat{R}_g$.
The minimum of this potential occurs for 
$\rho\approx 0.67$, to be compared with 
$\rho\approx 0.5$ of $V_{12}$ and $V_{23}$. 
Overlaps are more suppressed
since $V_{tr}(0) - V_{tr}(0.67) \approx 1.55k_BT$ 
(for our potentials we find $0.9k_BT$-$1.0k_BT$). Morever, the 
potential $V_{tr}(\rho)$ increases much less than $V_{12}(\rho)$ or 
$V_{23}(\rho)$ as $\rho$ increases. For instance, 
for $\rho \approx 1.5$, 
a value which occurs quite frequently, see figure~\ref{dist-radial.fig},
we have $V_{tr}(\rho) \approx 3.5k_BT$, while 
$V_{12}(\rho) \approx 4.9k_BT$, 
$V_{23}(\rho) \approx 6.1k_BT$. 

\subsection{Determination of the CGBM intermolecular potentials} \label{sec2.3}

As for the intermolecular potentials, we have made some drastic 
simplifications. First, we do not consider $n$-body interaction terms, which,
as we already mentioned, are important only for densities 
$\Phi\gtrsim n^{3\nu - 1}$.
Then, we consider a single intermolecular pair potential $W(\rho)$:
the atoms interact with the same potential, irrespective of their positions 
along the chains. Such a potential has been obtained by requiring 
the CGBM to reproduce the center-of-mass intermolecular distribution function.
Indeed, define in the polymer model
\begin{equation}
g_{CM}(r) = 
\langle e^{-\beta U_{12}} \rangle_{0,\bf r},
\end{equation}
where $\langle\cdot \rangle_{0,\bf r}$ indicates the average over two
polymers, the centers of mass of which are in the origin and in
$\bf r$, respectively, and $U_{12}$ is the intermolecular energy.
In the scaling limit $L\to\infty$, $g_{CM}(r)$ converges to 
a universal function $f_{CM}(\rho)$, $\rho = r/\hat{R}_g$. The pair potential 
has been determined so that 
\begin{equation}
g_{CM,CGBM}(\rho) = f_{CM}(\rho),
\label{eq-gCM}
\end{equation}
where $g_{CM,CGBM}(\rho)$ is the corresponding distribution function
in the CGBM.  Note that the second virial coefficient $B_2$ is 
related to $g_{CM}(r)$ by
\begin{equation}
B_2 = {1\over 2} \int d^3 {\bf r} [1 - g_{CM}(r)] = 
      2 \pi \int r^2 dr [1 - g_{CM}(r)].
\label{defB2}
\end{equation}
Hence, equality (\ref{eq-gCM}) guarantees that 
the adimensional combination $A_2 \equiv B_2/\hat{R}_g^3$, hence the 
thermodynamics in the small-density limit, agrees
in the CGBM and in the polymer model. 

\begin{figure}
\begin{center}
\begin{tabular}{c}
\epsfig{file=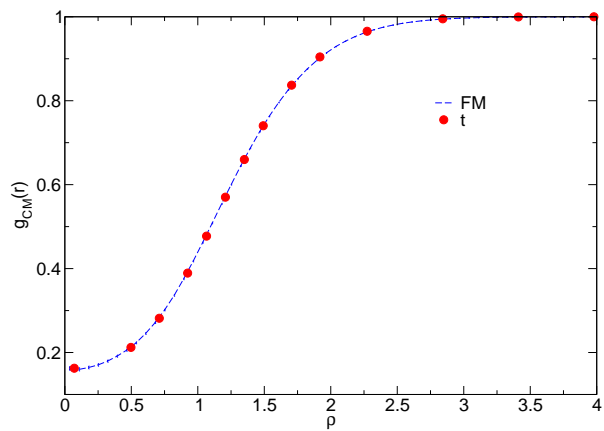,angle=-0,width=8truecm} 
    \hspace{0.5truecm} \\
\end{tabular}
\end{center}
\caption{Correlation function $g_{CM}(\rho)$ as a function
of $\rho = r/\hat{R}_g$. We report full-monomer results (dashed line)
and results (filled circles) 
for the model with potential (\ref{Wdef}).
}
\label{grcm0.fig}
\end{figure}

The potential $\beta W(\rho)$ has been parametrized as
\begin{equation}
\beta W(\rho) =c_1 \exp (-c_2 \rho^2), 
\label{Wdef}
\end{equation}
in terms of two unknown parameters $c_1$ and $c_2$. 
They have been determined following the approach of 
Akkermans {\em et al.}\cite{Akkermans:2001p1716,Akkermans:2001p6210}
Requiring the model to reproduce at best the polymer scaling
function $\rho^2 f_{CM}(\rho)$, we obtain the optimal values
$c_1 = 1.66$ and $c_2 = 3.9$.
For these parameter values the model with potential (\ref{Wdef})
has an intermolecular pair distribution function which agrees quite 
precisely with the corresponding polymer quantity, see 
figure~\ref{grcm0.fig}. The result depends on the parametrization and 
we cannot exclude that a different parametrization with the same 
number of parameters gives results of 
better quality.
Potential (\ref{Wdef}) 
differs only slightly from the intramolecular potential
$V_{14}(\rho)=1.86\exp(-4.08425\rho^2)$. 
Interactions between the atoms at the ends of the chain or between
atoms that belong to different chains are quite similar. 
It is interesting to compare our result
with that one would obtain 
by using the transferability hypothesis as suggested 
by Pierleoni {\em et al.}:
\cite{Pierleoni:2007p193} $\beta W_{tr}(\rho) = 1.92 \exp(-3.85 \rho^2)$.
The range of the potential is the same, but the potential we obtain is 
somewhat softer.

\section{Comparison of CGBM and polymer results} \label{sec.3}

In order to understand how well the tetramer model reproduces the 
polymer behavior we have performed extensive simulations 
of the tetramer and of a polymer model at zero density and at
volume fractions $\Phi = 1.09$, 4.36, 8.72. Since CGBM and polymer results
should be compared taking $\Phi = \Phi_b$, see 
(\ref{Phip-def}) and (\ref{Phib-def}), we drop the suffix
and thus $\Phi$ also refers to $\Phi_b$.
At finite density polymers have been 
modelled by means of the Domb-Joyce (DJ) model with $w =0.505838$, a value 
which is close to the optimal one for which no leading scaling corrections
are present (see Caracciolo {\em et al.}\cite{Caracciolo:2006p587} 
for details on the model). It allows us to determine precisely
the universal, model-independent scaling functions by using chains 
of moderate length. We consider walks of length $L=600$ and $L=2400$,
verifying the practical absence of scaling corrections. As in previous 
work\cite{Pelissetto:2008p1683}, we considered different box sizes, finding
negligible size effects when the number of chains in the box exceeds 100.
Simulations have been performed using the algorithm described in 
Pelissetto.\cite{Pelissetto:2008p1683}

\subsection{Zero-density}

\begin{figure}
\begin{center}
\begin{tabular}{c}
\epsfig{file=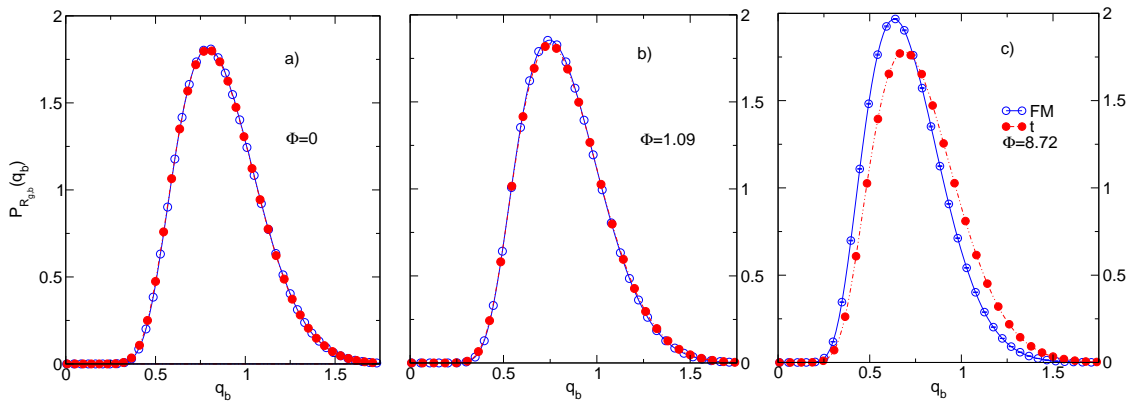,angle=-0,width=15truecm} 
    \hspace{0.5truecm} \\
\end{tabular}
\end{center}
\caption{Adimensional distribution $P_{R,b}(q_b)$ 
of the ratio $q_b = \hat{R}_{g,b}/\langle \hat{R}_g^2\rangle^{1/2}$, 
see definition (\protect\ref{distPRb}), for the tetramer (t)
and for polymers (FM). We report full-monomer (FM) and tetramer (t) results
for $\Phi = 0$ (a), $\Phi = 1.09$ (b) and $\Phi = 8.72$ (c).
}
\label{dist-Rgb-ZD}
\end{figure}

By construction, the tetramer CGBM reproduces the bond-length distributions.
As we have already remarked in Sec.~\ref{sec2.1}, see (\ref{Gintra-def}) and
(\ref{Rg-Gintra}), 
the ratio $\hat{R}_{g,b}/\hat{R}_g$ can be expressed in terms of these 
distributions, hence this ratio should assume the same value in the 
tetramer and in the polymer case.
Numerically, we find $\hat{R}_{g,b}/\hat{R}_g = 0.89093(7)$
for the tetramer and $\hat{R}_{g,b}/\hat{R}_g = 0.89210(11)$ for the 
polymer case. The difference is approximately 0.1\%, which shows 
how accurate the intramolecular potentials we determined are. Moreover, 
not only $\hat{R}_{g,b}/\hat{R}_g$ agrees, but also its
distribution (\ref{distPRb}) is the same for polymers and tetramers,
see figure~\ref{dist-Rgb-ZD}a). This is a nontrivial check, since this 
distribution is not directly related to the bond-length distributions,
nor to those of the bending and torsion angles.
Clearly, the tetramer models correctly
the shape and size of a polymer at zero density.
 
Since we have matched the center-of-mass distribution function to 
determine the intermolecular potential, the tetramer CGBM should 
give the correct second virial coefficient. If we expand the compressibility
factor as 
\begin{equation}
Z = {\Pi\over k_B T c} = 1 + B_2 c + B_3 c^2 + O(c^3),
\end{equation}
the quantity $A_2 = B_2/R_g^3$ is universal. An accurate estimate 
is \cite{Caracciolo:2006p587} $A_2 = 5.500(3)$. For the tetramer 
we obtain $A_{2,t} = 5.597(1)$. The difference is approximatey 1.8\%
and is representative of the level of precision with which the 
tetramer model reproduces the center-of-mass distribution  function. 
Much more interesting is the comparison of the third virial coefficient,
since it provides an indication of the accuracy with which the 
tetramer model reproduces the polymer thermodynamics in the dilute regime
and also of the importance of the neglected three-body forces.
The universal combination $A_3 = B_3/R_g^6$
was computed by Caracciolo {\em et al.}\cite{Caracciolo:2006p587} finding
$A_3 = 9.80(2)$.

In order to determine $A_3$, two contributions had to be computed. One 
contribution is the standard one, the only present in monoatomic 
fluids and in fluids of rigid molecules, $A'_3 \approx 10.64$, 
while the second one is 
a flexibility contribution $A_{3,fl} \approx -0.84$ (it corresponds to 
$-T_1 \hat{R}_g^{-6}$ in the notations of 
Caracciolo {\em et al.}\cite{Caracciolo:2006p587}). 
The combination $A_3$ as well as the two contributions 
$A'_3$ and $A_{3,fl}$ are universal, hence it makes sense to compare them with
the tetramer corresponding results. We obtain
\begin{equation} 
A_{3,t} = 9.99(2), \qquad 
A'_{3,t} = 10.57(2), \qquad
A_{3,fl,t} = -0.581(5).
\end{equation}
The tetramer reproduces quite reasonably the third virial coefficient,
the difference being approximately 2\%. Note that much of the 
discrepancy is due to $A_{3,fl}$: the tetramer is more rigid than the polymer.
It is interesting to compare these results with those obtained by using the 
single-blob model in which polymers are represented by monoatomic 
molecules interacting by means of density-independent potentials.
\footnote{If we were using density-dependent potentials, the 
thermodynamics, hence all virial coefficients, would be exactly reproduced.
However, here we only consider models with density-independent
potentials, hence the only meaningful comparison is with the single-blob model 
in which the interactions are density independent.}
If we use the accurate pair potential of 
Pelissetto {\em et al.}\cite{Pelissetto:2005p296} we 
obtain $A_3 = 7.844(6)$ (of course here $A_{3,fl} = 0$) which deviates
by 20\% from the polymer result. Hence, the tetramer model represents 
a significant improvement for the thermodynamics.

\begin{figure}
\begin{center}
\begin{tabular}{c}
\epsfig{file=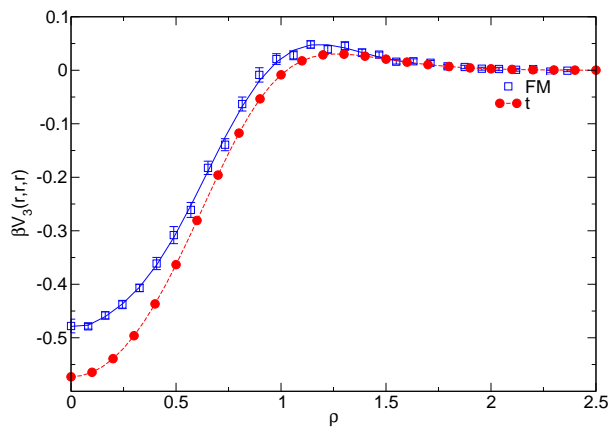,angle=-0,width=8truecm} \hspace{0.5truecm} \\
\end{tabular}
\end{center}
\caption{Three-body potential of mean force 
$\beta V_3({\bf r}_{12},{\bf r}_{13},{\bf r}_{23})$ for 
${r}_{12}={r}_{13} = {r}_{23} = r$, as a function of 
$\rho = r/\hat{R}_g$. We report the tetramer (t) result and the prediction
of full-monomer (FM) simulations. 
}
\label{threebody}
\end{figure}

As a further check we compute the effective three-body potential of 
mean force defined by \cite{Bolhuis:2001p288,Pelissetto:2005p296}
\begin{equation}
\beta V_3({\bf r}_{12},{\bf r}_{13},{\bf r}_{23}) = 
- \ln {\langle e^{-\beta U_{12} - \beta U_{13} - \beta U_{23}} 
    \rangle_{{\bf r}_{12},{\bf r}_{13},{\bf r}_{23}} \over 
    \langle e^{-\beta U_{12}}\rangle_{{\bf r}_{12}}
    \langle e^{-\beta U_{13}}\rangle_{{\bf r}_{13}}
    \langle e^{-\beta U_{23}}\rangle_{{\bf r}_{23}}  };
\end{equation}
here $U_{ij}$ is the intermolecular potential energy 
between tetramers $i$ and $j$
and the average 
$\langle \cdot \rangle_{{\bf r}_{12},{\bf r}_{13},{\bf r}_{23}}$
is over triplets of tetramers such that 
${\bf r}_{ij} = {\bf r}_{i} - {\bf r}_{j}$, where 
${\bf r}_{i}$ is the position of the center of mass of tetramer $i$.

We computed $\beta V_3({\bf r}_{12},{\bf r}_{13},{\bf r}_{23})$ 
for equilateral triangular configurations 
such that ${r}_{12}={r}_{13} = {r}_{23} = r$. The result is reported
in figure~\ref{threebody} and compared with the analogous quantity
computed in full-monomer simulations. 
At variance with the single-blob  model for which $\beta V_3 = 0$,
the tetramer model reproduces the polymer $\beta V_3$ quite reasonably: 
differences --- the tetramer potential is slightly more 
attractive --- are observed for 
$\rho = r/\hat{R}_g\lesssim 1$, but they are only significant for 
$\rho \lesssim 0.5$, i.e., when the tetramers are very close.
This is consistent with the analysis of the third virial coefficient:
in the dilute limit three-body interactions are correctly reproduced
by the tetramer model, without the need of introducing a three-body 
potential among tetramer atoms.

\subsection{Semidilute regime}

\begin{figure}
\begin{center}
\begin{tabular}{c}
\epsfig{file=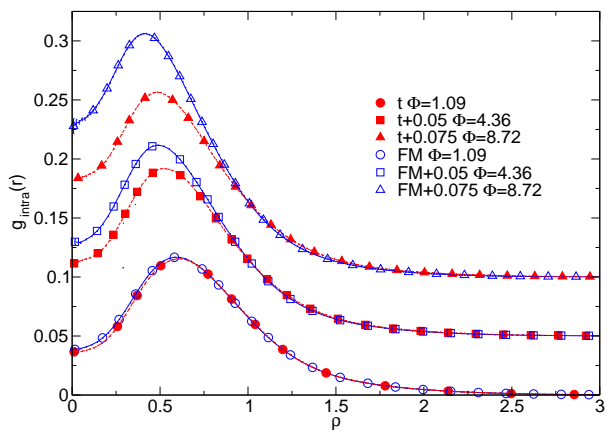,angle=-0,width=8truecm} \\
\end{tabular}
\end{center}
%% \vspace{1cm}
\caption{Adimensional intramolecular distribution function $g_{\rm intra}(r)$
as a function of $\rho = r/\hat{R}_g$. 
We report tetramer (dashed lines, t) and full-monomer (full lines, FM)
results. Distributions corresponding to $\Phi = 4.36$  and $\Phi = 8.72$ are 
shifted upward for clarity according to the legend.
}
\label{gintra}
\end{figure}

As we have discussed in the introduction, the tetramer model is expected to 
reproduce the full-monomer results up to $\Phi\approx 2$, representing a 
significant improvement with respect to the single-blob model which 
shows large deviations already for $\Phi = 1$. To check this expected behavior
 we compare tetramer and full-monomer simulations at 
$\Phi = 1.09$, 4.36, and 8.72.

Let us begin with the structural properties. In figure~\ref{gintra}
we report the adimensional intramolecular distribution function
$g_{\rm intra}(r)$. For $\Phi = 1.09$ the agreement between 
the tetramer and the full monomer results is excellent. 
However, as $\Phi$ increases, deviations
are observed for $\rho = r/\hat{R}_g\lesssim 1$. 
For $\Phi\gtrsim 4$ the tetramer is more 
swollen than the polymer: the probability for two blobs to be 
at a given distance $\rho\lesssim 1$
is significantly smaller in the tetramer than in the full-monomer chain. 
These results are 
further confirmed by the results for the radius of gyration. 
For the tetramer we have 
\begin{equation}
{R_{g,b}(\Phi)\over \hat{R}_g} = 
  \cases{
      0.85636(4)    &     $\qquad \Phi = 1.09,$ \cr
      0.8181(2)     &     $\qquad \Phi = 4.36,$ \cr
      0.8047(1)     &     $\qquad \Phi = 8.72,$ }
\end{equation}
to be compared with the full-monomer results 
\begin{equation}
{R_{g,b}(\Phi)\over \hat{R}_g} = 
  \cases{
      0.8523(2)     &     $\qquad \Phi = 1.09,$ \cr
      0.7823(2)     &     $\qquad \Phi = 4.36,$ \cr
      0.7346(6)     &     $\qquad \Phi = 8.72.$ }
\end{equation}
For $\Phi = 1.09$ the agreement is very good, consistently with 
the results reported in figure~\ref{gintra}. As $\Phi$ increases, however,
the tetramer is more rigid than the polymer and $R_{g,b}(\Phi)/\hat{R}_g$ 
is larger in the tetramer than in the polymer case. The same conclusions 
are reached by looking at the distribution of $R_{g,b}$, 
see figure~\ref{dist-Rgb-ZD}. For $\Phi = 1.09$ the agreement is excellent,
while for $\Phi = 8.72$ the tetramer distribution is slightly shifted 
towards larger values of $R_{g,b}$.

It is also interesting to compare the results for the bending and torsion 
angles reported in figure~\ref{angoli.fig}. 
The distributions appear to have a tiny 
dependence on $\Phi$ and to be reasonably reproduced 
by the tetramer for all values of $\Phi$. For instance, for the largest 
value of $\Phi$, $\Phi = 8.72$, we have for polymers
$P_b(\cos\beta = -1) \approx 0.93$, $P_t(\theta = 0) \approx 0.346$,
to be compared with 0.88 and 0.354, respectively, in the tetramer case.

\begin{table}
\caption{Compressibility factor for the tetramer model, $Z_t(\Phi)$,
and for polymers, $Z_p(\Phi)$. Polymer results are taken from
Pelissetto \protect\cite{Pelissetto:2008p1683}.
}
\label{Z-tetramer}
\begin{center}
\begin{tabular}{ccc}
\hline\hline
$\Phi$ & $Z_t(\Phi)$ & $Z_p(\Phi)$\\
\hline
0.054 &  1.07363(3) & 1.0725 \\
0.135 &  1.18993(4) & 1.1871 \\
0.27  &  1.39852(6) & 1.3929 \\
0.54  &  1.8499(1)  & 1.8536 \\
1.09  &  2.9090(1)  & 2.9589 \\
2.18  &  5.2660(2)  & 5.6342 \\
4.35  &  10.2056(4) & 12.229 \\
6.53  &  15.2279(1) & 20.019 \\
8.72  &  20.2811(1) & 28.716 \\
\hline\hline
\end{tabular}
\end{center}
\end{table}

\begin{figure}
\begin{center}
\begin{tabular}{c}
\epsfig{file=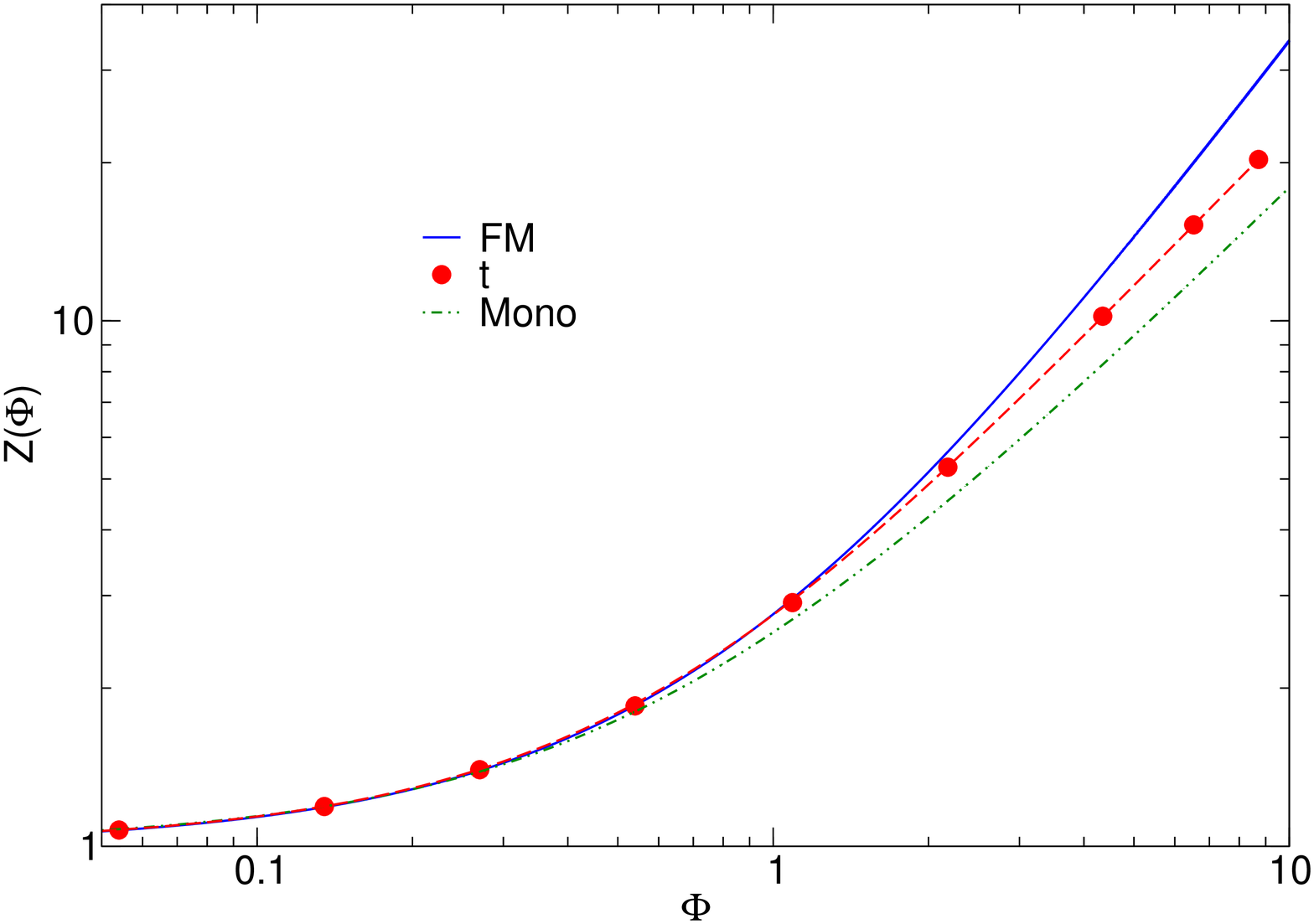,angle=-0,width=8truecm}  \\
\end{tabular}
\end{center}
%%\vspace{1cm}
\caption{Compressibility factor for polymers (FM), for the 
tetramer (t) and for the single-blob (mono) coarse-grained model.
For the tetramer we also show the interpolation
$Z(\Phi) = {(1+a_1 \Phi +a_2 \Phi^2+ a_3\Phi^3)^{1/2}/
      (1+a_4 \Phi)^{1/2} }$,
with $a_1 = 3.32676$, $a_2 = 4.67289$, $a_3 = 3.58551$, 
$a_4 = 0.65439$, of the data  reported in Table~\protect\ref{Z-tetramer}.
}
\label{fig:eostetra}
\end{figure}

Let us now consider the thermodynamics. For this purpose we computed
the compressibility factor $Z = \beta\Pi/c$
for the tetramer using the molecular virial 
method \cite{Ciccotti:1986p2263,Akkermans:2004p2261}
($c=N/V$ is the number concentration). As $Z$ is 
dimensionless, polymer and tetramer results at the same value of $\Phi$ 
can be directly compared. Estimates are reported 
in Table~\ref{Z-tetramer}.
For $\Phi \lesssim 1$ the tetramer $Z$ is very close to the 
polymer prediction: for $\Phi = 1.09$ it differs by 2\% from the correct result.
As $\Phi$ increases, however, differences increase and the 
tetramer model underestimates the correct pressure. 
In figure~\ref{fig:eostetra} we compare $Z(\phi)$ for the 
tetramer with the corresponding 
expression for polymers.\cite{Pelissetto:2008p1683} At the scale of 
the figure, good agreement is observed up to $\Phi \approx 2$. 
For larger densities, $Z(\Phi)$ in the tetramer increases slower than 
in the polymer case. Indeed, while in polymers we expect 
$Z\sim \Phi^{1/(3\nu-1)}\sim \Phi^{1.31}$ for large $\Phi$, for the 
tetramer $Z$ is expected to increase only linearly with $\Phi$
(since the potential is soft, for $\Phi \to \infty$ the random-phase
approximation should become exact \cite{Louis:2000p289}).

As can be seen from figure~\ref{fig:eostetra},
the tetramer model 
is significantly better than the single-blob model,\footnote{
For the single-blob model we shall always use the accurate expression
of the pair potential given in Pelissetto 
{\em et al.}\cite{Pelissetto:2005p296}} in which each 
polymer is represented by a single atom. At $\Phi = 1.09$ 
such a model gives $Z = 2.70(1)$, which underestimates $Z$
by 8\%, much more than the tetramer model (at this density the 
error on $Z$ is 2\%). 
The single-blob model gives a value for $Z$ 
which differs from the polymer one by less than 2\% only up to
$\Phi \approx 0.38$, i.e. up to densities which are a factor-of-three 
smaller than the corrisponding ones for the tetramer. This improvement 
confirms the scaling argument we presented in the introduction. 
As we explained in the introduction,
the multiblob model should give the correct thermodynamics up to 
a polymer volume fraction $\Phi_{\rm max}$ which scales as $n^{3\nu-1}$. 
If we compare the tetramer model with the single-blob one, we thus 
expect the density range in which the model is predictive 
to increase by $4^{3\nu-1}\approx 2.9$, which is 
indeed what we find.

\begin{figure}
\begin{center}
\begin{tabular}{c}
\epsfig{file=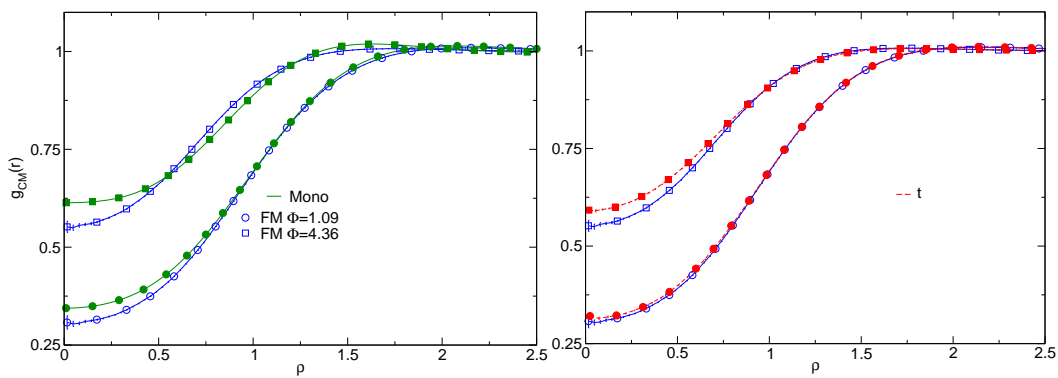,angle=-0,width=14truecm}  \\
\end{tabular}
\end{center}
%%\vspace{1cm}
\caption{Center-of-mass distribution function $g_{CM}(\rho)$ as 
a function of $\rho = r/\hat{R}_g$ for polymers (FM, open symbols)
(obtained from full-monomer simulations),
for the single-blob coarse-grained model (Mono, left, closed symbols) and for the tetramer 
(t, right, closed symbols).
We report results for $\Phi = 1.09$ and $\Phi = 4.36$.  
}
\label{fig:gr}
\end{figure}

Let us finally compare the center-of-mass distribution
function $g_{CM}(\rho)$. It is reported in figure~\ref{fig:gr} for $\Phi = 1.09$
and $\Phi = 4.36$. In the first case, the tetramer result is on top 
of the polymer results. For $\Phi = 4.36$ small discrepancies 
at short distance ($\rho\lesssim 0.5$) are present.
For instance, for the tetramer we have 
$g_{CM}(0) = 0.591(6)$, $g_{CM}(0.5) = 0.688(1)$ at $\Phi = 4.36$,
respectively,
to be compared with the polymer results 
$g_{CM}(0) = 0.550(5)$, $g_{CM}(0.5) = 0.660(2)$. 
These small differences are responsible for the differences in the 
estimates of $Z$
since $g_{CM}$ is related to $Z$ by the compressibility 
rule\footnote{In principle one can use this expression to determine 
the density derivative of $cZ$.
However, since the relevant length scale for $g_{CM}(\rho)$ is the 
average distance $d$ between the centers of mass of the polymer chains,
finite-size effects will be small only if $d/M\ll 1$, where $M$ 
is the size of the box containing the system. On the other hand, if 
one uses the intermolecular structure factor, the relevant scale is 
the correlation length $\xi$, which, in the semidilute regime, is 
significantly smaller than $d$. Therefore, $\xi/M$ is smaller than $d/M$,
which implies that determinations using the intermolecular structure factor
show smaller finite-size effects than those using $g_{CM}(\rho)$. }
\begin{equation}
\left({\partial cZ\over \partial c}\right)^{-1} = 
1 + c \int (g_{CM}(r)-1) d^3{\bf r}.
\end{equation}
Note that, even though the thermodynamics is poorly reproduced,
also the single-blob model gives an estimate of $g_{CM}(r)$ 
which is only slightly different from the polymer one. 
The largest differences are observed for $\rho\to 0$.
At overlap we obtain $g_{CM}(0) = 0.344(1)$ and $g_{CM}(0) = 0.613(1)$ for 
$\Phi = 1.09$ and $\Phi = 4.36$, to be compared with the 
polymer results $0.307(9)$ and $0.550(5)$.

\section{Comparison with other models} \label{sec.4}

In the previous section, we discussed the finite-density behavior of the 
tetramer model and we showed that it is quite accurate, for both 
structural and thermodynamic properties, up to $\Phi \approx 2$, 
in agreement with the multiblob argument of Pierleoni {\em et al.}
\cite {Pierleoni:2007p193} presented in the introduction.
It represents a significant improvement with respect to the single-blob model, 
which 
is unable to reproduce structural properties and reproduces the thermodynamics
only deep in the dilute regime (the compressibility factor $Z$ 
for the single-blob model differs from the polymer one
by less than 5\% only for $\Phi\lesssim 0.75$).

Here we wish to investigate the structural and thermodynamic behavior 
of two other models discussed in the literature. 

\subsection{Definition of the models} 

First, we consider the model introduced by Pierleoni {\em et al.}
\cite{Pierleoni:2007p193} ---
we name it model M1.
A CGBM with $n$ blobs is a chain in which neighboring atoms 
belonging to the same chain interact with an intramolecular potential 
\begin{equation}
V_{\rm bond}(r) = A e^{-\alpha r^2/\hat{r}_g^2} + k (r/\hat{r}_g - r_0)^2;
\label{Vbond-M}
\end{equation}
atoms that belong to the same chain but are not neighbors, or 
belong to different chains interact with a potential given by 
\begin{equation}
V_{\rm non-bond}(r) = A e^{-\alpha r^2/\hat{r}_g^2},
\label{Vnonbond-M}
\end{equation}
where $A$ and $\alpha$ are the same as in (\ref{Vbond-M}).
In these expressions $\hat{r}_g$ is the 
average zero-density radius of gyration of the blobs and 
sets the length scale. 
The model depends on several constants, which can be easily
determined in the dimer case, i.e., for $n=2$ 
(see caption of figure~1 in Pierleoni {\em et al.}\cite{Pierleoni:2007p193}):
$A = 1.92$, $\alpha = 0.80$, $k = 0.534$, and $r_0 = 0.730$.
To extend the model to values $n > 2$, Pierleoni {\em et al.}
\cite{Pierleoni:2007p193}
made the transferability hypothesis: equations
(\ref{Vbond-M}) and (\ref{Vnonbond-M})  hold for any $n$, the $n$-dependence
being completely taken into account by the radius of gyration of the blob.

As discussed in Sec.~\ref{sec2.2}, when comparing the CGBM results to the 
polymer ones, one should use the radius of gyration $\hat{R}_g$ of the 
reference polymer model and not the radius of gyration $\hat{R}_{g,b}$ of the 
CGBM. The radius $\hat{R}_g$ (or rather the ratio $\hat{R}_g/\hat{r}_g$, 
since $\hat{r}_g$ is the basic length scale in this approach) 
can be determined by using two different routes. 
As suggested by Pierleoni {\em et al.},\cite{Pierleoni:2007p193} 
one can determine 
$\hat{R}_{g,b}/\hat{r}_g$ for the CGBM and then use (\ref{Rg-Rgb}).
Alternatively one can use (\ref{scaling-rg2}).
If the model were a good CGBM,
these two definitions would give the same result, and indeed in the 
tetramer case they do. Instead, for model M1 we observe quite 
large differences. For instance, for $n=30$, we find 
$\hat{R}_g^2 \approx 41.8 \hat{r}_g^2$ if we use 
$\hat{R}_g^2 = \hat{R}_{g,b}^2 + \hat{r}_g^2$ 
and $\hat{R}_g^2 \approx 51.4 \hat{r}_g^2$, if we 
use $\hat{R}_g^2/ \hat{r}_g^2 = n^{2\nu}/1.06$.
These differences do not disappear as $n\to \infty$. 
An analysis of M1 results with $n\le 600$ gives for $n\to \infty$ the 
scaling behavior
\begin{equation}
{\hat{R}_{g,b}^2\over \hat{r}_g^2} = A n^{2\nu}, \qquad A = 0.78(3),
\label{scalingRgb-M1}
\end{equation}
which is not compatible with (\ref{scaling-rg2}).

In this paper we compare the results obtained by using
three different ``polymer" radii of gyration:
\begin{eqnarray}
\hat{R}_{g,1}/\hat{r}_g &=&   \hat{R}_{g,b}/\hat{r}_g  ,
  \label{Rg1}\\
\hat{R}_{g,2}/\hat{r}_g &=&   \sqrt{\hat{R}^2_{g,b}/\hat{r}^2_g + 1}  ,
  \label{Rg2} \\
\hat{R}_{g,3}/\hat{r}_g &=&   n^\nu/\sqrt{1.06}  .
  \label{Rg3}
\end{eqnarray}
Note that, for large $n$, definitions $\hat{R}_{g,1}$ and $\hat{R}_{g,2}$ 
are equivalent. On the other hand, as we discussed, 
definition $\hat{R}_{g,3}$ differs significantly from the others for any $n$,
including the limit $n\to\infty$. 
Recently, Coluzza {\em et al.}
\cite{Coluzza:2011p1723} suggested that model M1 should not be 
considered as a CGBM, but rather as a generic polymer model in 
good-solvent conditions,
so that $\hat{R}_{g,b}$ should be used as reference scale. In the 
following we shall mainly focus on the first and third definition
and we shall label the corresponding results as (M1,1) and (M1,3),
respectively.

A proper definition of $\hat{R}_g$ is relevant for two purposes: 
first, length distributions are universal only if one 
expresses the lengths in terms of $\hat{R}_g$ (the relevant scale is 
$\rho = r/\hat{R}_g$); second, at finite density results should be compared 
at the same value of the polymer volume fraction $\Phi$ defined 
in (\ref{Phip-def}).
Changing the definition of $\hat{R}_g$ changes the definition of both
$\rho$ and $\Phi$, hence it is crucial to specify which $\hat{R}_g$ one is 
using.  Note that Coluzza {\em et al.} \cite{Coluzza:2011p1723} 
introduced a complex procedure to compare 
CGBM and polymer results. 
Their procedure is fully equivalent to the one we have 
discussed above: to analyze finite-density results, 
one must compare the results at the same value of the 
adimensional volume fraction $\Phi$.

The thermodynamic behavior of model M1 
was studied by Pelissetto.\cite{Pelissetto:2009p287}
If $\hat{R}_{g,b}$ is used as reference scale
as recently suggested by Coluzza {\em et al.},\cite{Coluzza:2011p1723}
the model fails to reproduce the 
thermodynamics unless $n$ is larger than $10^3$, but of course, for such
values of $n$, there are several other models --- the lattice Domb-Joyce model
we use is one of them --- which better reproduce the universal polymer
behavior both for 
the thermodynamics and the structural properties.
For instance, for $n = 100$, which is a relatively large value, 
model M1 overestimates the 
second virial coefficient combination $A_2$ by 20\%.
A more detailed discussion will be presented below.

We shall also consider a second coarse-grained model, 
\cite{Pelissetto:2009p287} we call it model M2. 
Conceptually, this was not intended to be 
a CGBM, but rather a polymeric model which reproduces
the asymptotic (number of monomers $n\to \infty$) behavior already for small 
values of $n$. The $n$-dependent potentials were tuned so that 
thermodynamics was reproduced for $\Phi \lesssim 10$. 
For $n=26$, thermodynamics was reproduced taking potentials of the 
form (\ref{Vbond-M}) and (\ref{Vnonbond-M}) with 
($\hat{r}_g$ is no longer the blob size, but simply sets the length scale)
$A = 8.28$, $\alpha = 1$, $k = 0.15$ and $r_0 = 0.653$.
It is important to stress that 
$\hat{R}_{g,b}$ was used as reference length
in the optimization procedure employed to 
determine the optimal parameters. 
Therefore, for consistency, for this model it makes no sense to 
use definitions $\hat{R}_{g,2}$  and $\hat{R}_{g,3}$. 
Hence, whenever we consider model 
M2, $\hat{R}_g$ should always be identified with $\hat{R}_{g,b}$.

We have performed simulations for model M1 for $n = 8$, 16, 30, 60 and 
of model M2 for $n = 30$.  In the second case, 
one should in principle compute the appropriate parameters for $n=30$.
We will use here the coefficients computed for $n=26$,
as we expect the changes 
necessary as $n$ goes from 26 to 30 to be tiny. 

\subsection{Numerical results and discussion}

\begin{figure}
\begin{center}
\begin{tabular}{c}
\epsfig{file=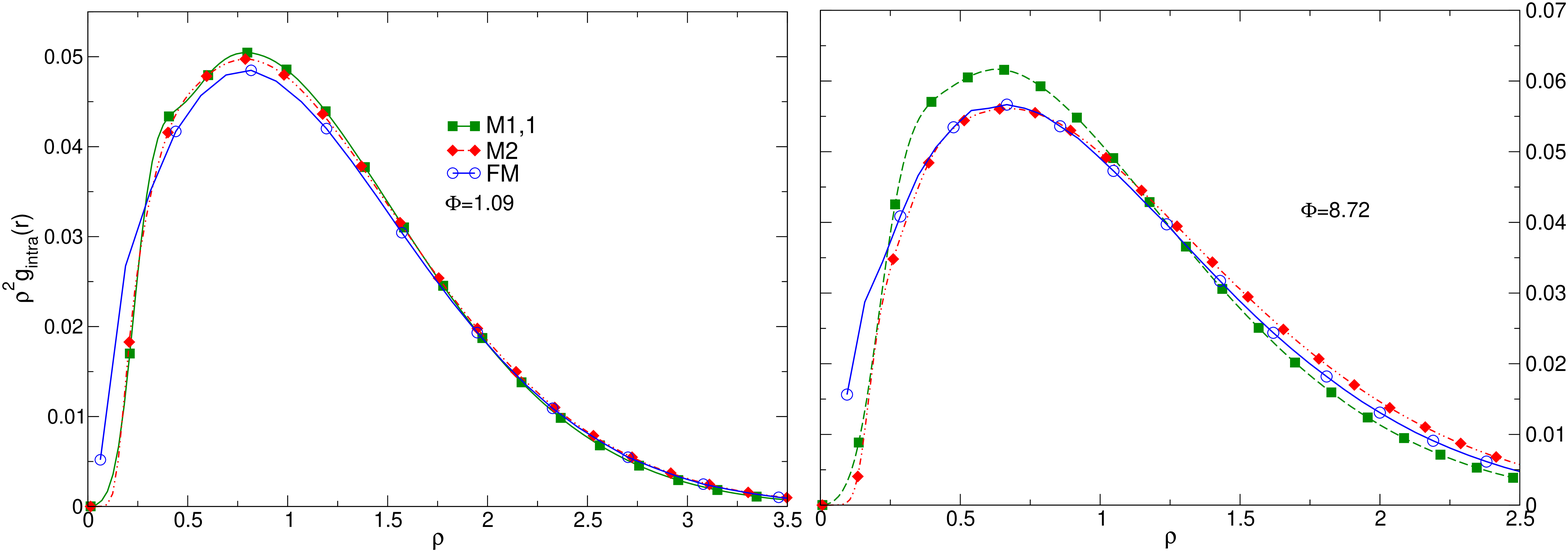,angle=0,width=12truecm} \hspace{0.5truecm} \\
\end{tabular}
\end{center}
\caption{Intramolecular distribution function 
$\rho^2 g_{\rm intra}(r)$ versus $\rho = r/\hat{R}_g$ for 
$\Phi = 1.09$ (left) and $\Phi = 8.72$ (right). We report full-monomer 
results (FM) and results for models M1 and M2 with $n=30$.
In models M1 and M2 (but not in the polymer case) 
we use $\hat{R}_{g,b}$ as radius of gyration, 
both in the definition of $\rho$ and in that of $\Phi$.
}
\label{gintra-mod}
\end{figure}

\begin{figure}
\begin{center}
\begin{tabular}{c}
\epsfig{file=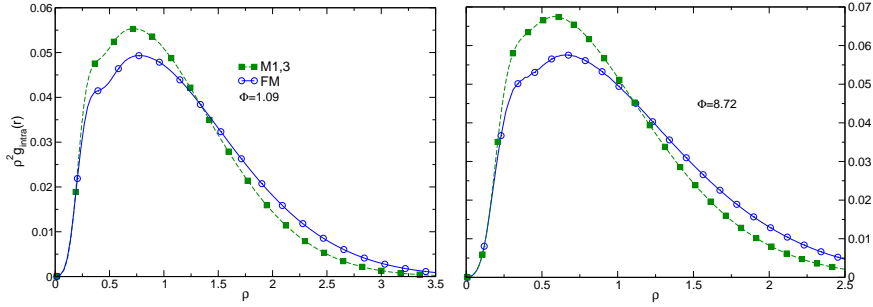,angle=0,width=12truecm} \hspace{0.5truecm} \\
\end{tabular}
\end{center}
\caption{Intramolecular distribution function 
$\rho^2 g_{\rm intra}(r)$ versus $\rho = r/\hat{R}_g$ for 
$\Phi = 1.09$ (left) and $\Phi = 8.72$ (right). We report full-monomer 
results for a 30-blob representation of the polymer (FM) 
and results for model M1 with $n=30$.
In model M1 we use $\hat{R}_{g,3}$, see definition (\ref{Rg3}), 
as radius of gyration. 
}
\label{gintra-mod-2}
\end{figure}

Let us first discuss the structural properties, considering the 
intramolecular distribution function $g_{\rm intra}(r)$. If we consider 
models M1 and M2 as generic polymer models --- in this case we should 
use $\hat{R}_{g,b}$ as length scale --- the corresponding results should 
be compared with the monomer intramolecular distribution function, 
which is defined in (\ref{gintra-def}), taking $n = L$.
Estimates of $\rho^2 g_{\rm intra}(\rho)$, which 
represents the average distribution of the bond lengths, 
are shown in figure~\ref{gintra-mod}.
At $\Phi = 1.09$,
we observe a reasonable agreement for both models: 
they appear to be able to reproduce the structural properties 
in the dilute regime. For $\Phi = 8.72$ model M1 shows 
some, but still reasonably small, differences for 
$0.4\lesssim \rho \lesssim 1$. Model M2 appears to be slightly in
better agreement, except for small $\rho \lesssim 0.2$.

\begin{figure}
\begin{center}
\begin{tabular}{cc}
\epsfig{file=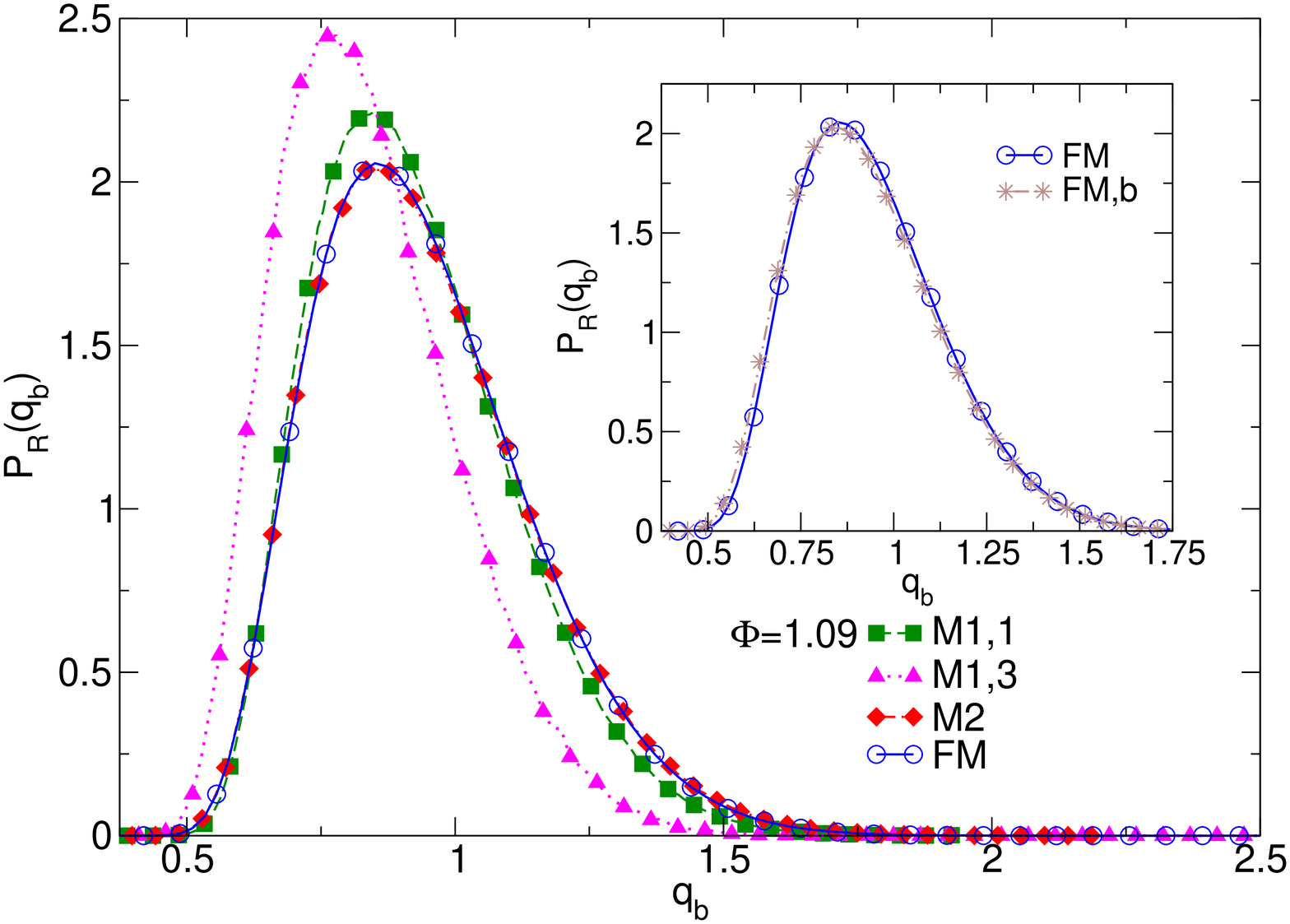,angle=0,width=7truecm} \hspace{0.5truecm} &
\epsfig{file=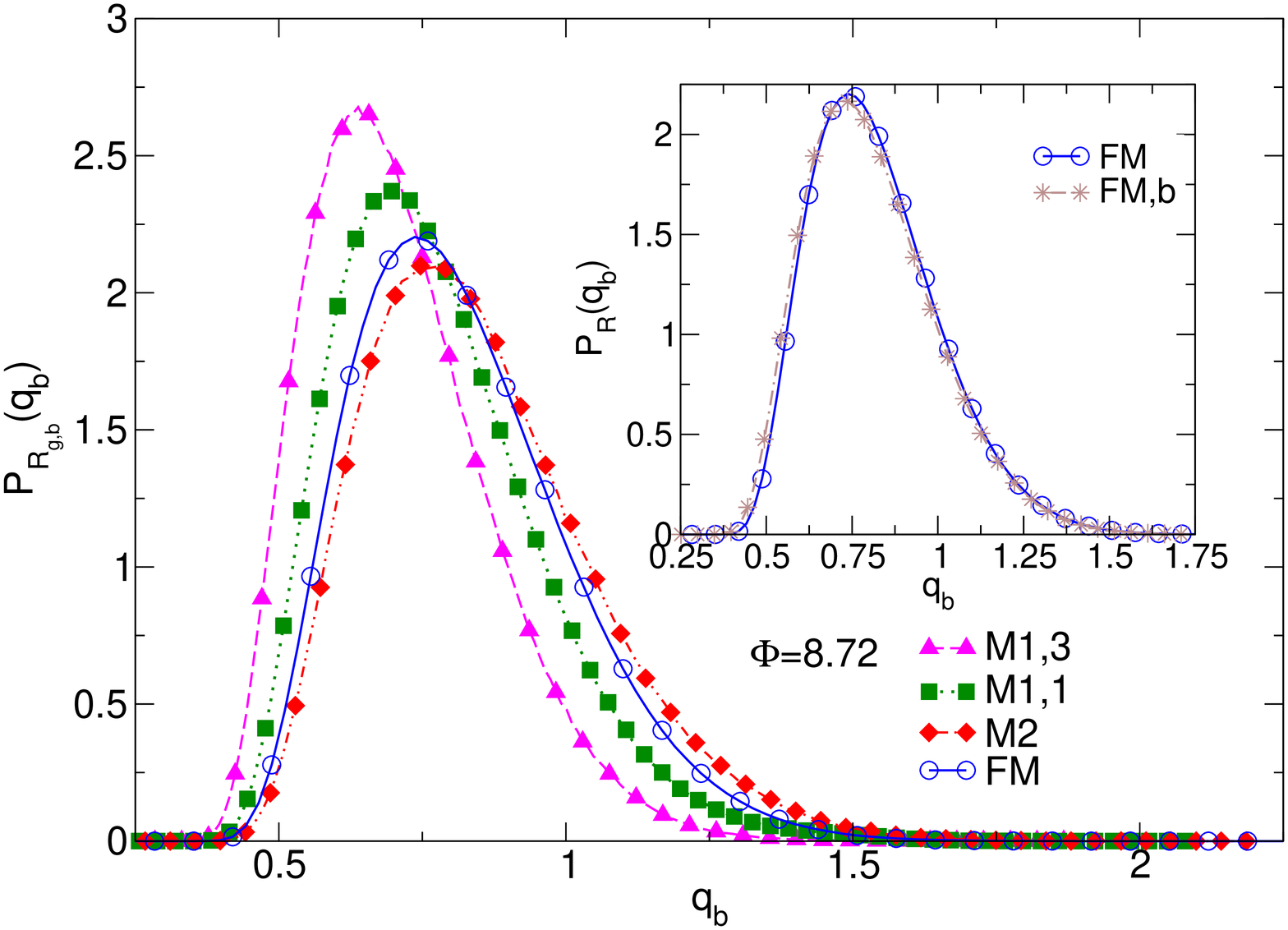,angle=0,width=7truecm} \hspace{0.5truecm} \\
\end{tabular}
\end{center}
\caption{Adimensional distribution $P_{R}(q_b)$ of the ratio $q_b$.
We report:  full-monomer 
results for $q_b = R_g/\langle \hat{R}_{g}^2 \rangle^{1/2}$ (FM);
results for model M2 with $n=30$, defining $\Phi$ using $\hat{R}_{g,b}$, with 
$q_b = R_{g,b}/\langle \hat{R}_{g,b}^2 \rangle^{1/2}$;
results for model M1 with $n=30$. In this last case we show two different
quantities: in case (M1,1) we use the same definitions as for model M2,
while in case (M1,3), we set 
$q_b = R_{g,b}/\langle \hat{R}_{g,3}^2 \rangle^{1/2}$
and define $\Phi$ in terms of $\hat{R}_{g,3}$.
In the inset we report again the FM data together with the 
distribution of $q_b$ defined in (\protect\ref{distPRb}),
as appropriate for a 30-blob system (FM,b).
}
\label{Rgdist-mod}
\end{figure}

\begin{figure}
\begin{tabular}{cc}
\epsfig{file=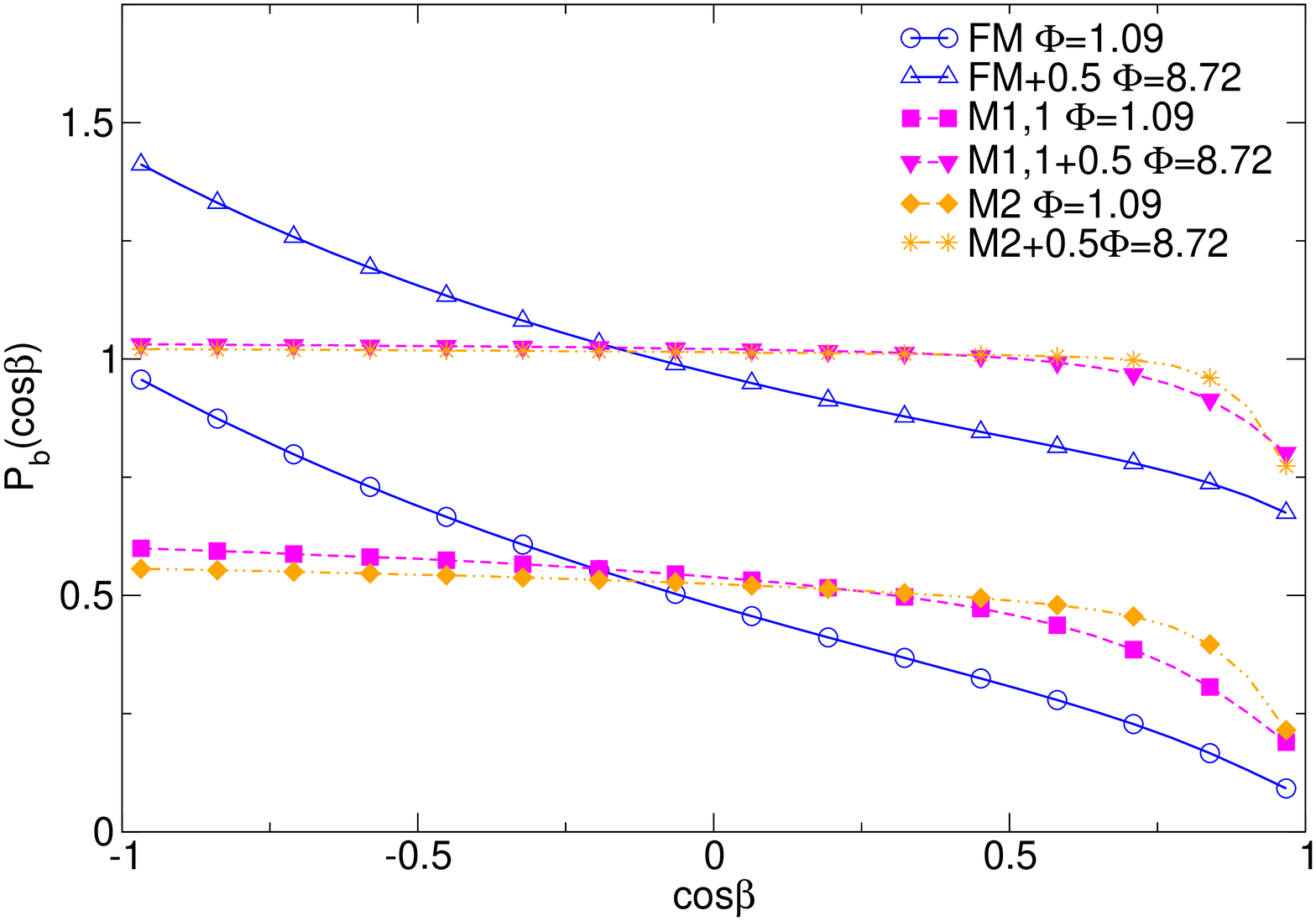,angle=0,width=7truecm} \hspace{0.5truecm} &
\epsfig{file=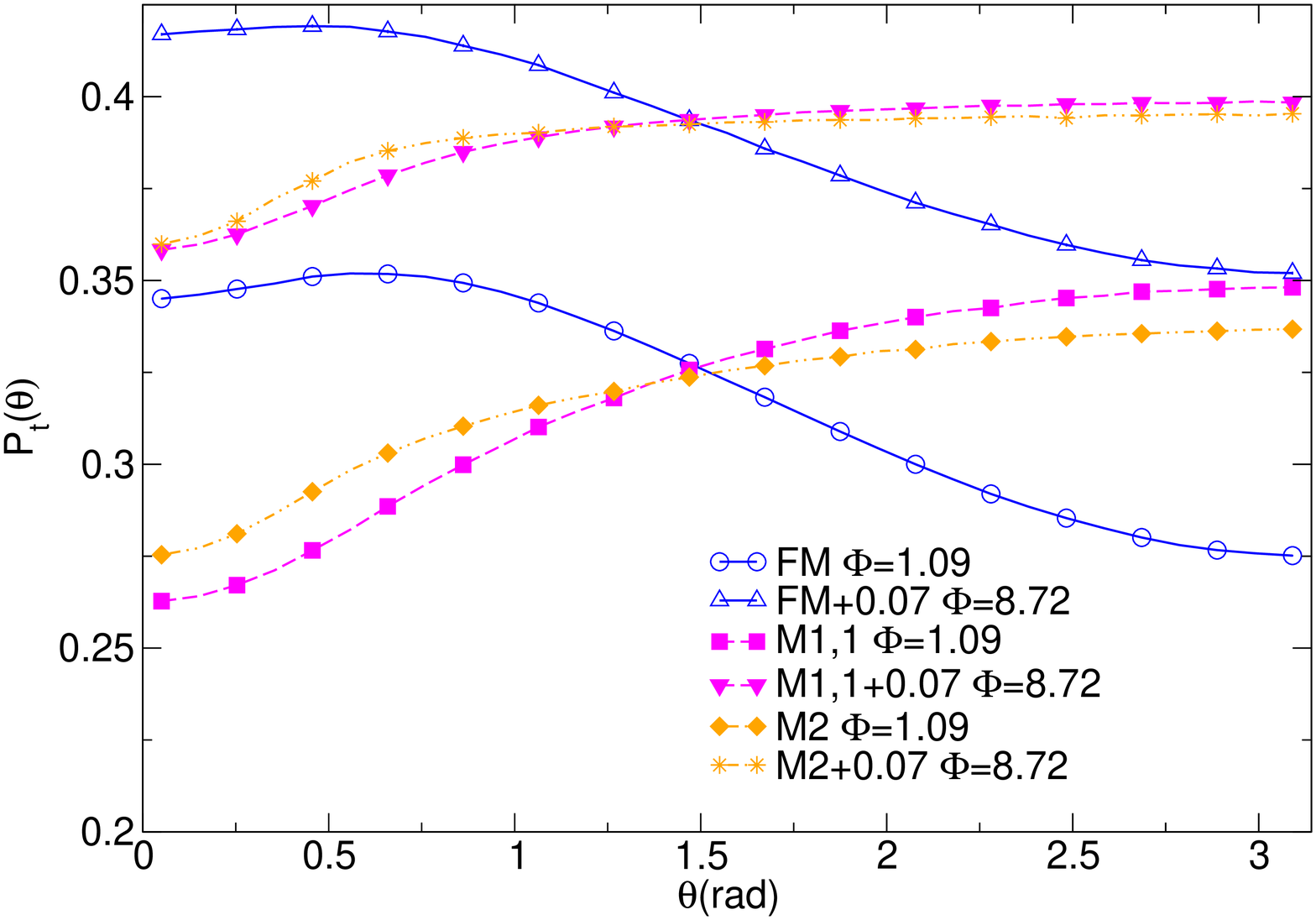,angle=0,width=7truecm} \hspace{0.5truecm} \\
\end{tabular}
\caption{Bending (left) and torsion (right) angle distributions.
We report full-monomer 
results for a 30-blob representation of the polymer (FM) 
and results for models M1 and M2 with $n=30$.
For M1 and M2,
the polymer volume fraction $\Phi$ is computed by using $\hat{R}_{g,b}$.
Distributions for $\Phi = 8.72$ are shifted for clarity.
}
\label{angles-mod}
\end{figure}

For model M1, we can also consider $\hat{R}_{g,3}$, 
see definition (\ref{Rg3}), for the zero-density radius of
gyration. In doing this, we implicitly assume that model M1 is a CGBM, 
as the tetramer model, and not just a generic good-solvent polymer model.
In figure~\ref{gintra-mod-2} we report the corresponding
adimensional intramolecular distribution function, which should 
be compared in this case with the polymer results for a 30-blob coarse-grained 
representation (data labelled FM).
Discrepancies are significantly larger than in figure \ref{gintra-mod}.
Clearly, structural properties are much better reproduced if $\hat{R}_{g,b}$ is 
used as radius of gyration, in agreement with the conclusions of 
Coluzza {\em et al.}
\cite{Coluzza:2011p1723} Note that if one uses $\hat{R}_{g,b}$ the M1 
distributions agree exactly with the polymer ones for $n\to \infty$. 
Indeed, model M1 is, in this limit, a generic polymer model and all models have 
the same infinite-length behavior as long as the same adimensional 
quantities are compared. 
As a consequence, the discrepancies we observe in figure~\ref{gintra-mod-2}
do not decrease as $n$ increases. 
Similar conclusions are reached by considering the distribution 
of the radius of gyration, see figure~\ref{Rgdist-mod}. 
Depending on the interpretation of the models as generic polymer models
or as CGBMs, 
one should compare the results with the polymer distributions of 
$R_g/\<\hat{R}_g\>^{1/2}$ or of $R_{g,b}/\<\hat{R}_g\>^{1/2}$,
where $R_{g,b}$ is the radius of gyration of 
a 30-blob representation of the polymer chain.
However, as shown in the inset of figure~\ref{Rgdist-mod},
the two distributions are identical on the scale of the figure, 
so that this conceptual difference is not relevant in practice.
Model M2 appears to be the one which gives the best agreement, 
but, if $\hat{R}_{g,b}$ is used as a reference scale, 
also the model-M1 distribution is close to the full-monomer one.
If instead $\hat{R}_{g,3}$ is used for model M1, discrepancies are quite large.

If model M1 is considered as a CGBM, it makes also sense
to compare the bending and torsion angle distributions. 
The results, reported in 
figure~\ref{angles-mod}, (similar to those observed in model M2) 
have little relation with what is observed for the polymer case 
(the full-monomer distributions we report are those appropriate for a 
30-blob representation of the polymer). 
Hence, even if the bond-length
distributions are approximately reproduced, correlations between 
different bonds, for instance angular distributions, are not,
and the true polymer shape
is quite different from that predicted by model M1.

\begin{table}
\caption{Estimates of $A_2 = B_2 \hat{R}_g^3$ 
and $A_3 = B_3 \hat{R}_g^6$ using the different definitions of 
$\hat{R}_g$ for model M1. Numerically, 
we find $\hat{R}^2_{g,b}/\hat{r}^2_g = 7.987(3)$,
18.82(1), 40.83(4), 95.37(3), for $n= 8$, 16, 30, 60. 
The universal asymptotic values for polymers are 
\protect\cite{Caracciolo:2006p587}
$A_2 = 5.500(3)$, $A_3 = 9.80(2)$. }
\label{table_A2A3_M1}
\begin{center}
\begin{tabular}{cccccccc}
\hline\hline
$n$   &   \multicolumn{3}{c}{$A_2$}  & \multicolumn{3}{c}{$A_3$} \\
      &  $\hat{R}_{g,b}$   & $\hat{R}_{g,2}$  & $\hat{R}_{g,3}$  &
       $\hat{R}_{g,b}$   & $\hat{R}_{g,2}$  & $\hat{R}_{g,3}$  \\
\hline
8  &     9.225(7)  &   7.729(5)  &   5.815(1)  &
         32.0(5)   &   22.4(3)   &   12.7(1) 
\\
16 &     8.258(9)  &   7.640(8)  &   5.548(1)  &
         26.0(5)   &   22.2(4)   &   11.7(1)   
\\
30 &     7.55(1)   &   7.28(1)   &   5.354(2)  &
         21.4(7)   &   20.0(7)   &   10.8(2)
\\
60 &     6.95(1)   &   6.84(1)   &   5.183(6)  & 
         17.9(8)   &   17.3(7)   &   10.0(4)   
\\
\hline\hline
\end{tabular}
\end{center}
\end{table}

\begin{figure}
\begin{center}
\begin{tabular}{c}
\epsfig{file=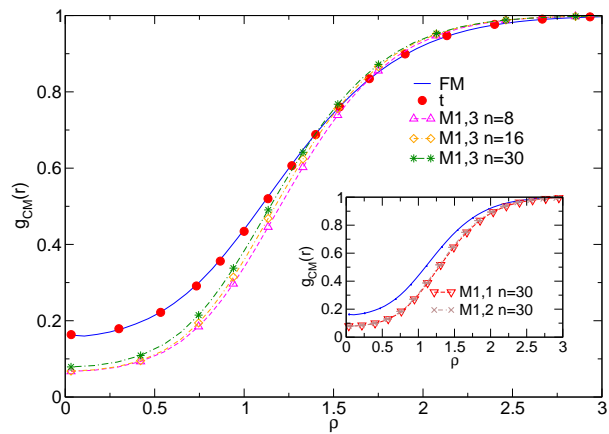,angle=0,width=8truecm} \hspace{0.5truecm} \\
\end{tabular}
\end{center}
\caption{
Center-of-mass pair distribution function $g_{CM}(r)$ as a function
of $\rho = r/\hat{R}_{g}$ for polymers (FM), tetramers (t), 
and model M1 for $n=8,16,30$
at zero density. For model M1, 
$\hat{R}_g^2 = \hat{R}_{g,3}^2 = n^{2\nu}/1.06$ in the
main panel, while in the inset we report results ($n=30$ only) with 
$\hat{R}_g = \hat{R}_{g,b}$ (M1,1) and $\hat{R}_g = \hat{R}_{g,2}$  
[see definition (\ref{Rg2})] (M1,2).
For polymers and tetramers, $R_g$ is always the radius of gyration.
}
\label{grcm0-mod}
\end{figure}

\begin{figure}
\begin{center}
\begin{tabular}{cc}
\epsfig{file=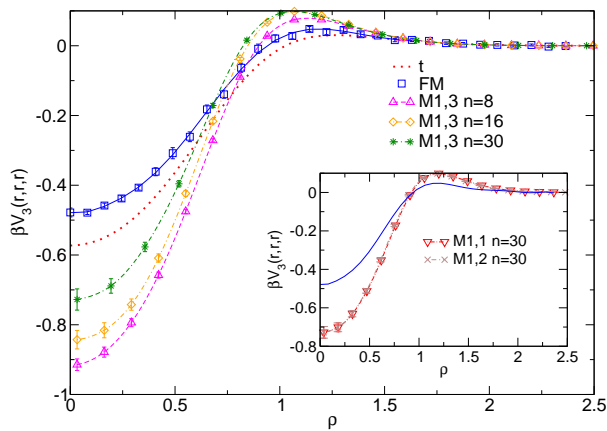,angle=0,width=8truecm} \hspace{0.5truecm} &
\end{tabular}
\end{center}
\caption{
Three-body potential of mean force $\beta V_3(r,r,r)$ for equilateral
triangular configurations as a function
of $\rho = r/\hat{R}_{g}$ for polymers (FM), tetramers (t),
and model M1 for $n=8,16,30$ at zero density.
For model M1, $\hat{R}_g = \hat{R}_{g,3}$ in the
main panel, while in the inset we report results with 
$\hat{R}_g = \hat{R}_{g,b}$ (M1,1) and $\hat{R}_g = \hat{R}_{g,2}$ 
[see definition (\ref{Rg2})] (M1,2).
For polymers and tetramers $\hat{R}_g$ is always the radius of gyration.
}
\label{gr3-mod}
\end{figure}

\begin{figure}
\begin{center}
\begin{tabular}{c}
\epsfig{file=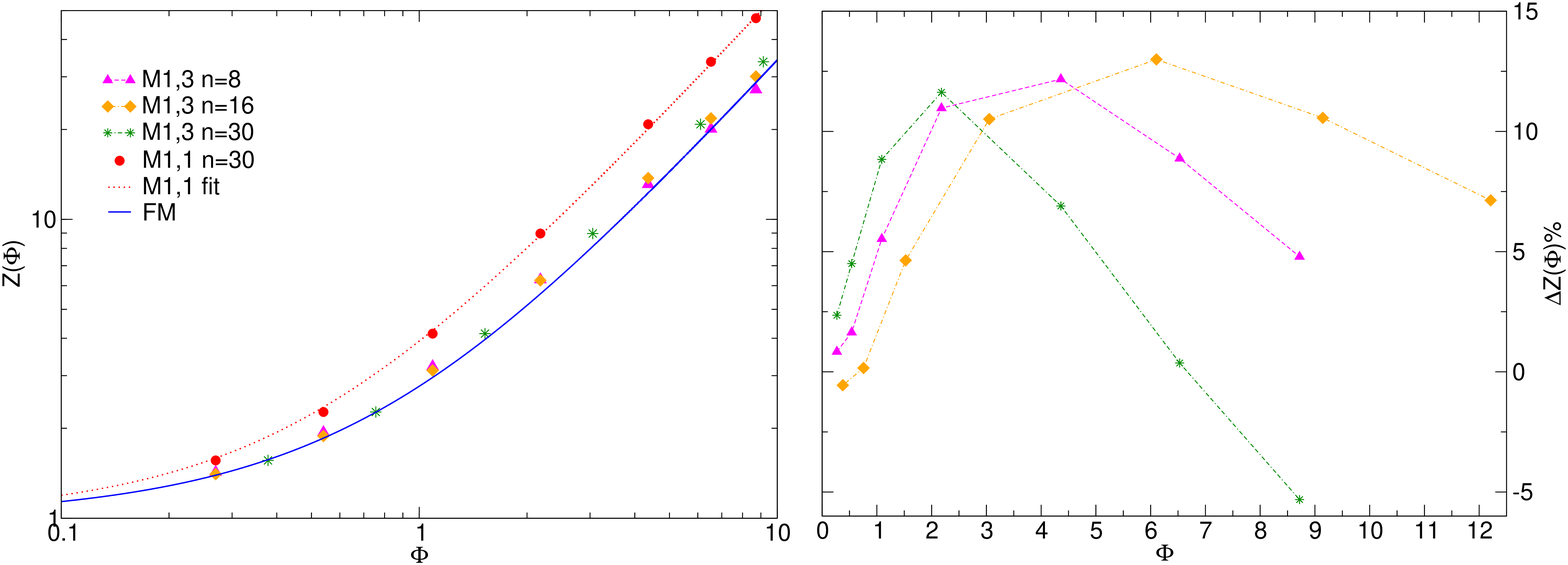,angle=0,width=14truecm} \hspace{0.5truecm}  \\
\end{tabular}
\end{center}
\caption{Left: compressibility factor $Z(\Phi)$ for polymers (FM) and 
model M1: in the latter case $\Phi$ is defined by using $\hat{R}_{g,b}$
(data labelled M1,1) or by using 
$\hat{R}_{g,3}$ (data labelled M1,3).
On the right we show the deviations 
of the (M1,3) results from the polymer ones, 
$\Delta Z = 100 [Z_{M1,3} (\Phi)/Z_{FM}(\Phi) - 1]$. 
}
\label{Z-mod}
\end{figure}

Let us now come to the thermodynamics. For both models we have determined
the second virial coefficient $B_2$ and the adimensional combination 
$A_2 = B_2 \hat{R}_g^{-3}$. The parameters of model M2 were determined 
in such a way to reproduce $A_2 = 5.500$, the correct result for infinitely
long polymers, hence M2 gives the correct thermodynamics in the zero-density
limit. The results for model M1 are reported in Table~\ref{table_A2A3_M1}
for each choice of $\hat{R}_g$. As already discussed 
by Pelissetto,
\cite{Pelissetto:2009p287} if $\hat{R}_{g,b}$ is used, $A_2$ differs
significantly from the asymptotic result, even for $n = 60$. If 
$\hat{R}_{g,2}$ is used, discrepancies are smaller for $n = 8$, but 
substantially the same for $n\ge 30$ (not surprising, since 
$\hat{R}_{g,2}/R_{g,b} \to 1$ as $n\to\infty$). Definition 
$\hat{R}_{g,3}$ gives apparently better results, but we believe that 
this apparent agreement is fortuitous. Indeed, as $n$ increases,
$B_2 \hat{R}_{g,3}^{-3}$ should monotonically decrease, increasing
the discrepancy with the polymer case. It is easy to compute the asymptotic 
value. For large $n$ model M1 is a generic good-solvent polymer model, 
hence standard universality arguments predict that $B_2 \hat{R}_{g,b}^{-3}$ 
should converge to 5.500, the result obtained for infinitely long polymers.
\cite{Caracciolo:2006p587}.  Using 
(\ref{scalingRgb-M1}) we obtain for $n\to \infty$
\begin{equation}
B_2 \hat{R}_{g,3}^{-3} = 
           (1.06 A)^{3/2} B_2 \hat{R}_{g,b}^{-3} = 
           {5.500 (1.06 \times 0.78)^{3/2}} \approx 4.13,
\end{equation}
which differs by 25\% from the correct result. To further confirm
that there is nothing fundamental in the observed agreement, we plot 
the zero-density center-of-mass distribution function $g_{CM}(\rho)$ 
in figure~\ref{grcm0-mod}. For all values of $n$ it differs significantly 
from the polymer one. In particular, the correlation hole $g_{CM}(0)$,
which does not depend on the choice of $\hat{R}_g$, is significantly 
deeper in model M1 than for good-solvent polymers in the scaling limit. 

In Table~\ref{table_A2A3_M1} we also report the third-virial combination
$A_3$. If $\hat{R}_{g,b}$ is used, results differ roughly by a factor
of two from the polymer ones. Discrepancies decrease if $\hat{R}_{g,3}$ is 
used, but again this is accidental. The same argument given above 
for $A_2$ shows that
the combination $B_3 \hat{R}_{g,3}^{-6}$ converges to 5.5
for $n\to\infty$, which is roughly a 
factor-of-two smaller than the correct result \cite{Caracciolo:2006p587} 
$A_3 = 9.80(2)$. We also
report, see figure~\ref{gr3-mod}, the three-body potential of mean force
for three chains on an equilateral triangle.
We observe significant discrepancies: results are significantly worse than
those obtained by using the tetramer CGBM.

\begin{figure}
\begin{center}
\begin{tabular}{c}
\epsfig{file=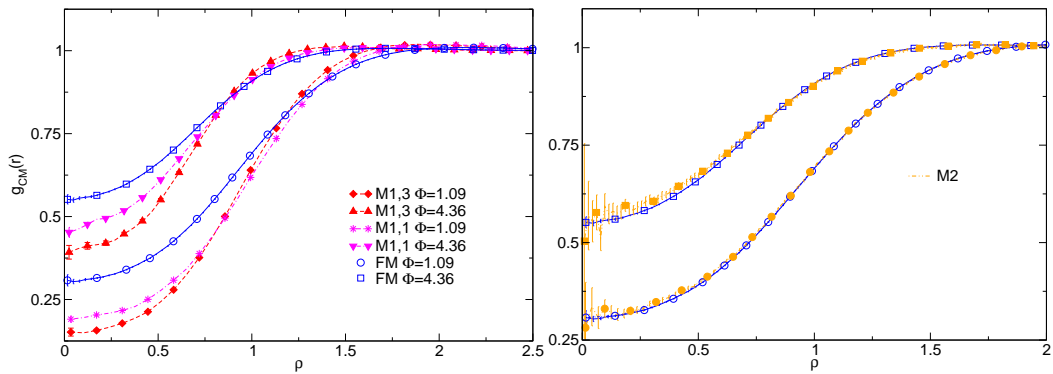,angle=0,width=14truecm} \hspace{0.5truecm}\\
\end{tabular}
\end{center}
\caption{Center-of-mass distribution function $g_{CM}(r)$ as a function 
of $\rho = r/\hat{R}_{g}$ for polymers (FM), 
model M1 (left) and model M2 (right), at 
densities $\Phi = 1.09$ and 4.36. The number of blobs is $n=30$.
For models M1 (case M1,1) and M2 (but not in the polymer case) 
we use $\hat{R}_{g,b}$ as radius of gyration, 
both in the definition of $\rho$ and in that of $\Phi$.
Data labelled (M1,3) are obtained by using $\hat{R}_{g,3}$ 
as radius of gyration,
both in the definition of $\rho$ and in that of $\Phi$.
}
\label{grcmPhi-mod}
\end{figure}

Let us now compare the thermodynamics at finite density. In figure~\ref{Z-mod}
we compare the compressibility factor for polymers (data labelled by FM
taken from Pelissetto \cite{Pelissetto:2008p1683})
with that for model M1. As expected on the basis of the zero-density results, 
if $\hat{R}_{g,b}$ is used in the definition of $\Phi$, very large 
discrepancies are observed. Moreover, also the dependence on $\Phi$ 
is incorrect: $Z(\Phi)$ increases as $\Phi^{1.13}$ 
for $6 \lesssim \rho \lesssim 9$, which differs significantly from
the correct scaling $\Phi^{1.31}$. Discrepancies are significantly 
smaller (12\% at most, see the right panel of figure~\ref{Z-mod}) 
if $\hat{R}_{g,3}$ is used. Again the agreement appears to be accidental,
since the center-of-mass distribution function differs significantly 
from the polymer one, see figure~\ref{grcmPhi-mod}. Even worse, 
for $\rho \lesssim 1$, results using $\hat{R}_{g,b}$ appear to be closer
to the correct full-monomer results than those obtained by using 
$\hat{R}_{g,3}$. Again, note that only if $R_{g,b}$ is used the distribution
function $g_{CM}(\rho)$ computed in model M1 will converge to the 
full-monomer one for $n\to\infty$. 
If $R_{g,3}$ is used instead, 
the correlation hole is always (even for $n\to\infty$) deeper for model M1 
than for true polymers at any given value of the polymer packing 
fraction $\Phi\not=0$. Moreover, $g_{CM}(\rho)$ shows more curvature,
reaching approximately 1 at a slightly smaller value of $\rho$.

By construction, model M2 reproduces the thermodynamics up to $\Phi = 10$.
Indeed, the parameters were fixed by requiring $Z(\Phi)$ to be equal to 
the polymer compressibility
in the dilute limit and for $\Phi = 10$. Note that it also
gives the correct intermolecular pair distribution function, see 
figure~\ref{grcmPhi-mod}, a result which is not {\em a priori} obvious.

\section{Conclusions} 

In the last two decades (but the first proposals \cite{FK-50} 
can be traced back to the '50s) 
several coarse-grained models have been proposed for polymers 
in solution under good- or $\theta$-solvent conditions. In the 
simplest approaches polymer chains are mapped onto single atoms 
interacting by means of soft potentials. These classes of models are however
unable to reproduce the structural properties and give the correct 
thermodynamics only in the dilute limit. To go to higher densities,
density-dependent potentials \cite{Louis:2000p269,Bolhuis:2001p268} 
may be used. However, their determination requires in any case 
finite-density full-monomer simulations, which is what one would like 
to avoid by using coarse-grained models. Moreover, it is not clear 
how accurate they are in more complex situations in which 
polymers only constitute one species in the solution. 
To overcome these difficulties, the multiblob approach was 
recently proposed,\cite{Pierleoni:2007p193} in which each polymer 
chain is mapped onto a short linear chain of $n$ blobs. This model
retains some degrees of freedom and thus it should allow us to 
obtain the correct thermodynamics even in the semidilute regime. 
The main difficulty of this approach is the derivation of the 
intramolecular interactions. In Pierleoni {\em et al.}\cite{Pierleoni:2007p193}
potentials 
were obtained for any value of $n$ on the basis of a transferability 
hypothesis.  However, later\cite{Pelissetto:2009p287} it was shown that 
the resulting model did not have the correct  thermodynamic behavior,
indicating that much more work was needed to determine the 
intramolecular interactions. 

In this paper we consider again the multiblob approach, 
determining the intramolecular interactions by matching 
universal zero-density polymer distributions.\footnote{
The polymer distributions are computed by using a lattice model.
However, standard renormalization-group arguments allow us to 
conclude that exactly the same results would be obtained in the 
limit $L\to\infty$ by using any other --- discrete or continuous --- model.}
We map polymer coils onto four-blob chains (tetramers) which interact 
be means of bonding, bending and torsional angle potentials. They are 
obtained by requiring the bond-length distributions and the 
distributions of the bending and torsion angles to be the same in 
the tetramer and in the full-monomer model at zero density. 
As for the intermolecular interactions, we only consider pairwise
blob-blob interactions, neglecting many-blob potentials.
This limits the validity of the 
model to the regime in which blob-blob overlaps are rare, i.e., to 
blob volume fractions $\eta_b = c_b/c_b^* \lesssim 1$ 
[$c_b$ is the blob concentration and $c_b^* = 3/(4 \pi \hat{r}_g^3)$].
For the tetramer this gives $\Phi\lesssim n^{3\nu-1}\approx 2.9$.

The tetramer model turns out to be quite accurate up to $\Phi\approx 2$, 
in agreement with the argument given above. In this range of densities
structural properties as well as the thermodynamics are correctly reproduced.
For instance, for $\Phi = 2.18$ the error on $Z(\Phi)$ is 7\%. 
If we compare the compressibility factor computed in the 
tetramer model to that determined in the single-blob model we observe a 
factor-of-three improvement, indicating that the ideas behind the 
multiblob approach really work. For $\Phi \gtrsim 2$ significant deviations
are observed, both for the structure --- tetramers are too rigid ---
and for the thermodynamics --- $Z(\Phi)$ in the tetramer model
becomes significanly smaller than for polymers as $\Phi$ increases.

We have investigated again the model proposed in \cite{Pierleoni:2007p193},
model M1, 
studying in detail structure and thermodynamics. We find that 
the model cannot be considered as a consistent CGBM, but should rather be 
thought as a generic polymer model, as recently suggested
by Coluzza {\em et al.}\cite{Coluzza:2011p1723} 
Since the thermodynamics is poorly 
reproduced for small values of $n$ (we mainly investigate the case 
$n=30$), it is not a surprise that for these numbers of blobs 
intermolecular correlations
are significantly different from those determined in full-monomer simulations
with a large number of monomers. On the other hand, internal bond 
distributions are quite well reproduced. Clearly, for small values of 
$n$, in spite of the poor thermodynamic behavior, model M1 is able to
model correctly some features of the 
polymer shape, though not all of them --- for instance, angle distributions
are not reproduced.  This is consistent with the results of 
Coluzza {\em et al.}\cite{Coluzza:2011p1723} 
They studied the geometric structure of polymer 
brushes, comparing results obtained in full-monomer simulations
and in model M1. Also in that case, good agreement was observed for 
some structural properties. 

Finally, we consider the model proposed by Pelissetto.\cite{Pelissetto:2009p287}
In this case, parameters were tuned so that the thermodynamics was exactly
reproduced up to $\Phi = 10$. We find that it also reproduces well the 
intermolecular structure: the polymer center-of-mass distribution function
is correctly reproduced in the whole density range $\Phi \lesssim 10$.
As for the intramolecular structure, we find that the model 
gives results analogous to those obtained for model M1. Bond-length 
distributions are approximately reproduced in the density range we have 
investigated, indicating that also this model correctly reproduces 
some features of the polymer shape.

In conclusion, we have shown that the newly proposed tetramer model is a
significant step forward in the development of a consistent coarse-grained 
model of polymer chains based on zero-density interactions.
To investigate the semidilute regime for large densities, i.e.,
for  $\Phi\gtrsim 2$, multiblob models with $n>4$
must be developed. In this respect, the most important lesson of the 
present work is that many-body intramolecular interactions cannot be completely 
neglected, if one aims at a consistent multiblob model; 
their absence in model M1 is probably the cause of its failure
in reproducing the thermodynamics of polymer solutions.
Finally, it would be very important --- we leave it for future work ---
to develop an analogous coarse-graining strategy for chains in 
$\theta$ conditions. Here single-blob models with pairwise 
intermolecular interactions fail since thermodynamic stability is only
obtained by taking into account
three-chain interactions. Since the tetramer model reproduces quite nicely 
three-chain correlations in the good-solvent regime, 
it is the good candidate to attack this problem. 

\section*{Acknowledgements} 

C.P. is supported by the Italian Institute of Technology (IIT) under the 
SEED project grant number 259 SIMBEDD – Advanced Computational Methods for 
Biophysics, Drug Design and Energy Research.

\appendix

\section{The radius of gyration of the blobs: universal predictions}
\label{App-rgblob}

In this appendix we wish to discuss the behavior of the radius of gyration
of the blobs. If $r_{g,i}(\Phi)$ is the radius of gyration of the $i$-th
blob along the chain, the ratio $r_{g,i}(\Phi)/R_g(\Phi)$ is universal, 
being an adimensional ratio of large-scale properties of the polymer.
It only depends on the position $i$ of the blob along the 
chain, on the number $n$ of blobs, and on the density through the polymer
volume fraction $\Phi$. 
Of course, this holds when the number of monomers $L$ is large, otherwise 
scaling corrections should be taken into account. In general we have
\begin{equation}
{r_{g,i}(\Phi,L,n) \over R_g(\Phi,L)} = 
   f_i(n,\Phi) \left(1 + k g_i(n,\Phi) L^{-\Delta} + \ldots \right),
\label{roverR}
\end{equation}
where $f_i(n,\Phi)$ and $g_i(n,\Phi)$ are universal functions, 
$\Delta = 0.528(12)$, see Clisby,\cite{Clisby:2010p2249} 
is a universal exponent,
and $k$ a nonuniversal constant that does not depend on 
$i$, $n$, and $\Phi$, but only on the model. In the polymer model we use
at finite density, the Domb-Joyce model with $w = 0.505838$, 
the constant $k$ is approximately zero, so that corrections
decay with the next-to-leading exponent $\Delta_2 \approx 1$.

An approximate expression for the $n$-dependence of the function
$f_i(n,\Phi)$ which works well for $\Phi\ll 1$ is obtained as follows.
In the large-$L$ limit we have standard Flory scaling, 
$R_g = b L^\nu$ and $r_{g,i} = b' (L/n)^\nu$, 
with\cite{Clisby:2010p2249} $\nu = 0.587597(7)$.
Now assume that the blob shape and size is not influenced by the
neighboring blobs, so that the size of the blob is equal to that 
of a free polymer with the same number of monomers. 
We can thus approximate $b' \approx b$,
so that $r_{g,i}/R_g = n^{-\nu}$. This formula is of course not exact, 
since blob-blob interactions cannot be neglected. Still, as we now show,
it is reasonably accurate for $\Phi \ll 1$. 

\begin{figure}
\begin{tabular}{cc}
\epsfig{file=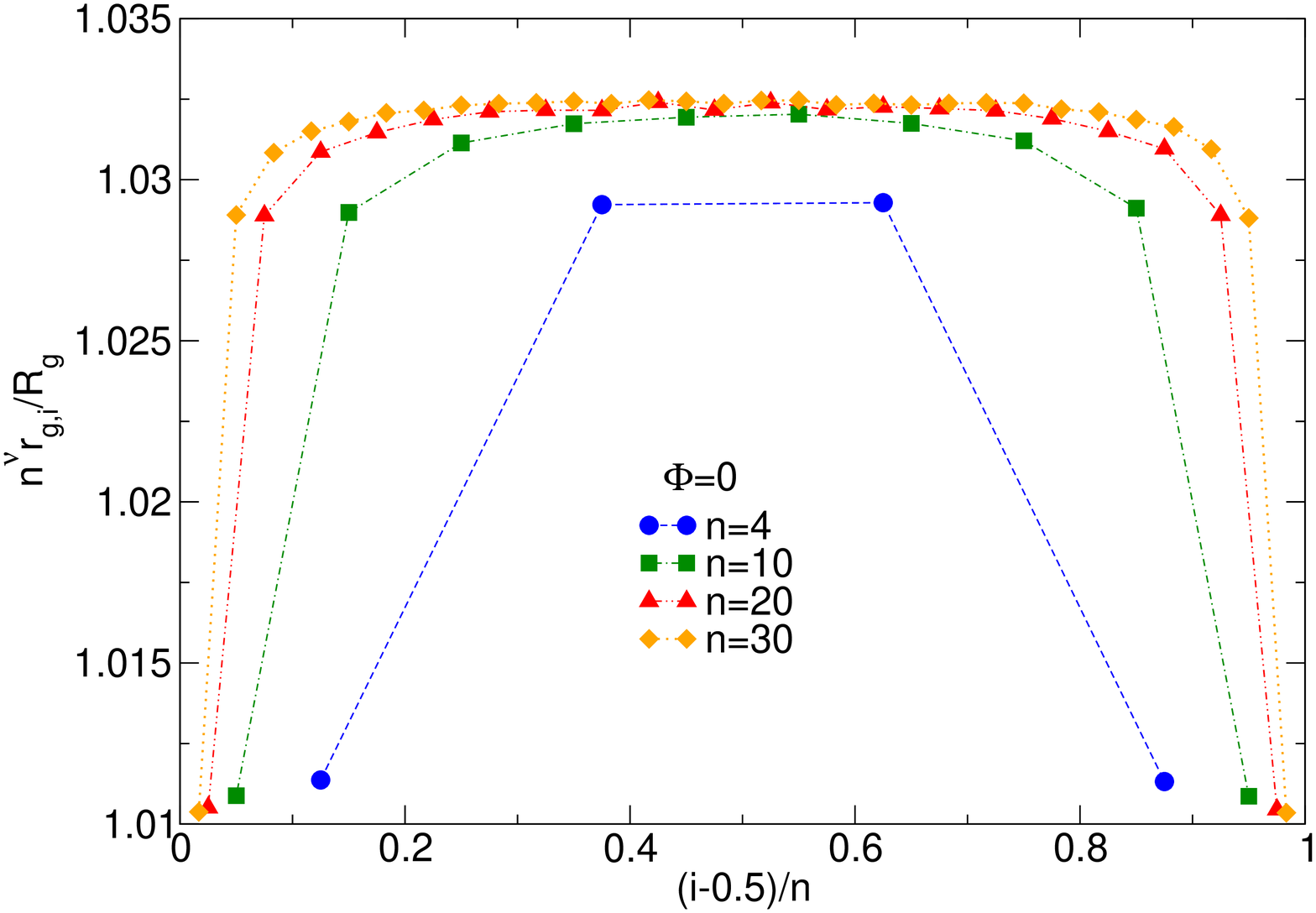,angle=0,width=7truecm} \hspace{0.5truecm} &
\epsfig{file=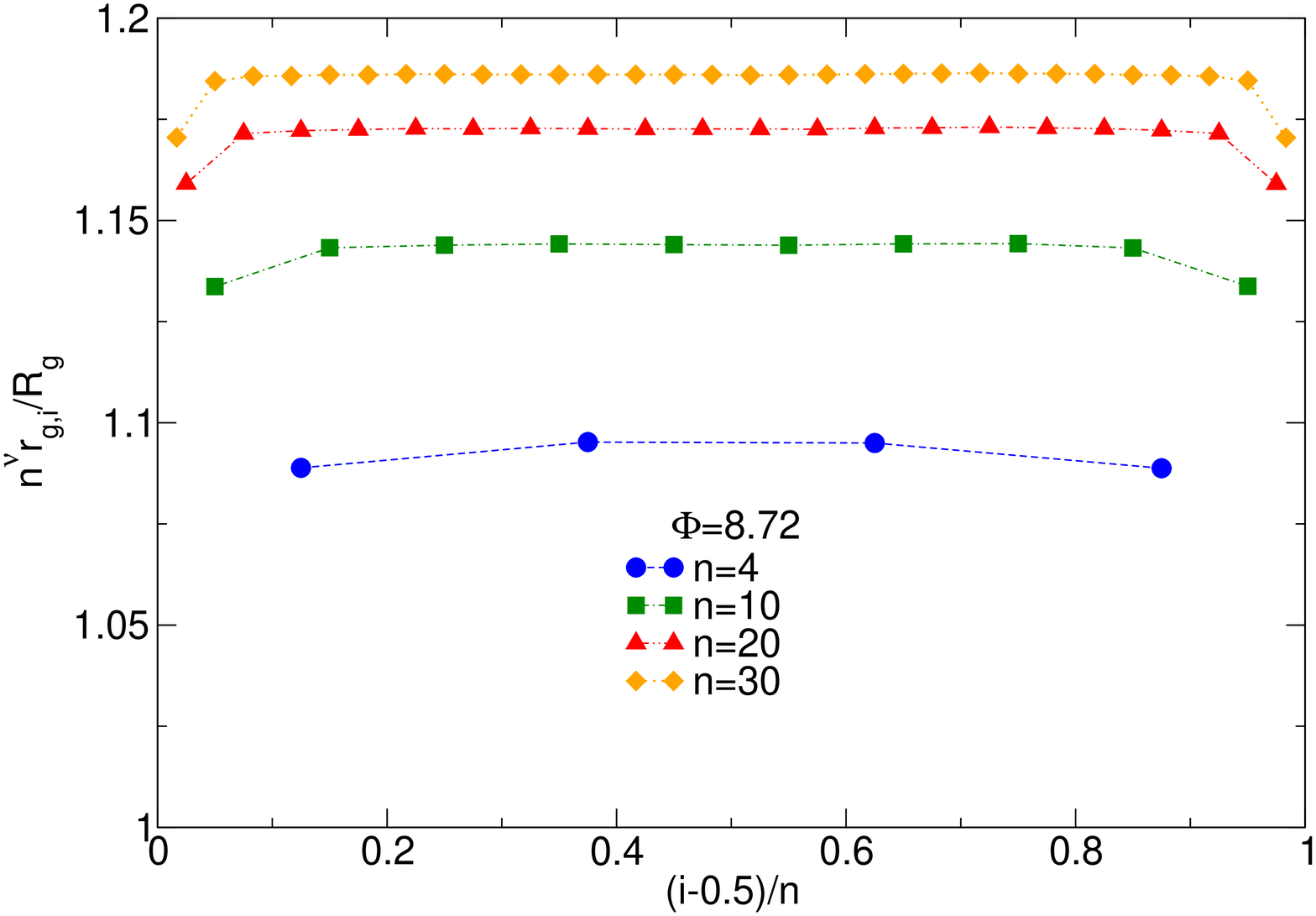,angle=0,width=7truecm} \hspace{0.5truecm} \\
\end{tabular}
\caption{Combination $n^\nu Q_{i} = n^\nu r_{g,i}/R_g$ as 
a function of $(i-1/2)/n$ for $n=4,10,20,30$. On the left we report
the results for $\Phi = 0$, on the right for $\Phi = 8.72$.
}
\label{Q-ratio}
\end{figure}

In order to compute $r_{g,i}/R_g$ in the asymptotic limit, 
we determine $Q_i(L,n) = r_{g,i}(n)/R_g$ for $L = L_1=600$ and $L = L_2 = 2400$
in the Domb-Joyce model with $w = 0.505838$. Assuming corrections with
exponent $\Delta_2 = 1.0(1)$, we estimate the asymptotic ($L\to\infty$)
value as 
\begin{equation}
   Q_{i,\rm as} (n)= 
   {L_1^{\Delta_2} Q_i(L_1,n) - L_2^{\Delta_2} Q_i(L_2,n) \over 
                   L_1^{\Delta_2} - L_2^{\Delta_2} }
\end{equation}
The combination $C(i,n,\Phi) = n^\nu Q_{i,\rm as}(n)$ for $\Phi = 0$ is 
reported in figure~\ref{Q-ratio} 
as a function of $(i-1/2)/n$ for several values of $n$. Note that 
this quantity is always larger than 1, indicating that a blob of  
$L/n$ monomers is more swollen than an isolated chain of the same degree
of polymerization.
This is due to the neighboring blobs which are entangled with the blob 
one is considering, causing swelling.
Second, this effect is smaller for the boundary blobs since they
have only one neighbor. 
The scaling $\hat{r}_{g,i}/\hat{R}_g  \sim n^{-\nu}$
holds quite well at zero density even for $n=4$, 
with a proportionality constant which is only slightly larger
than 1. In particular, for the boundary blobs we have 
$\hat{r}_{g,i}/\hat{R}_g  \sim   1.01 n^{-\nu}$, while for the internal blobs
$\hat{r}_{g,i}/\hat{R}_g  \sim   1.03 n^{-\nu}$. 
If we average over all blobs and 
neglect end effects, we obtain
relation (\ref{scaling-rg2}), which we used extensively in the text.

The swelling effect is expected to increase as $\Phi$ increases, since the
higher the density the higher the blob-blob entanglement is. 
In figure~\ref{Q-ratio}
we also report $C(i,n,\Phi)$ for $\Phi = 8.72$. There are here two notable 
differences with respect to the case $\Phi = 0$. First of all,
end effects are small, indicating that much of the swelling is due 
to neighboring chains, consistently with the idea that for $\Phi \gtrsim 1$ 
polymers are strongly intertwined. Second, the $n$ dependence of the 
scaling function $f_i(n,\Phi)$ defined in Eq.~(\ref{roverR}) is not 
captured by the simple scaling form $n^{-\nu}$ 
for our small values of $n$ (of course, $f_i(n,\Phi)$ scales as $n^{-\nu}$
for $n\to \infty$). 

Given the blob radii of gyration, using Eq.~(\ref{Rg-Rgb}), 
we can compute the ratio $R_{g,b}/R_g$.
For $n = 4$ we obtain 
\begin{equation}
   {R_{g,b}(\Phi)\over R_{g}(\Phi)}
     = \cases{
              0.89210(10)  &   $\qquad \Phi = 0$  \cr
              0.88701(10)  &   $\qquad \Phi = 1.09$ \cr
              0.87937(11)  &   $\qquad \Phi = 4.36$ \cr
              0.8753(4)    &   $\qquad \Phi = 8.72$  }
\label{Rgbratio-n=4}
\end{equation}
Note that the $\Phi$ dependence is tiny.
At $\Phi = 0$, a good approximation for all $n\ge 4$ is given by
\begin{equation}
   {\hat{R}_{g,b}\over \hat{R}_{g}} = 
    \sqrt{1 - k n^{-2\nu} } \qquad k = \left(1.03 - 0.04/n\right)^2,
\label{RgboverR-pred}
\end{equation}
which predicts for ${\hat{R}_{g,b}/\hat{R}_{g}}\approx 0.8922$ for $n=4$,
in good agreement with the result (\ref{Rgbratio-n=4}).

\begin{table}
\caption{Ratio $r_g(\Phi,n)/\hat{r}_g(n)$ as a function of 
$\Phi$ and $n$.}
\label{ratiorg}
\begin{center}
\begin{tabular}{ccccc}
\hline\hline
$\Phi$ & $n=4$ & $n=10$ & $n=20$ & $n=30$ \\
\hline
1.09 & 0.982   & 0.990  &  0.994 & 0.996  \\
4.36 & 0.938   & 0.962  &  0.976 & 0.982  \\
8.72 & 0.898   & 0.933  &  0.955 & 0.965  \\
\hline\hline
\end{tabular}
\end{center}
\end{table}

\begin{figure}
\begin{center}
\begin{tabular}{c}
\epsfig{file=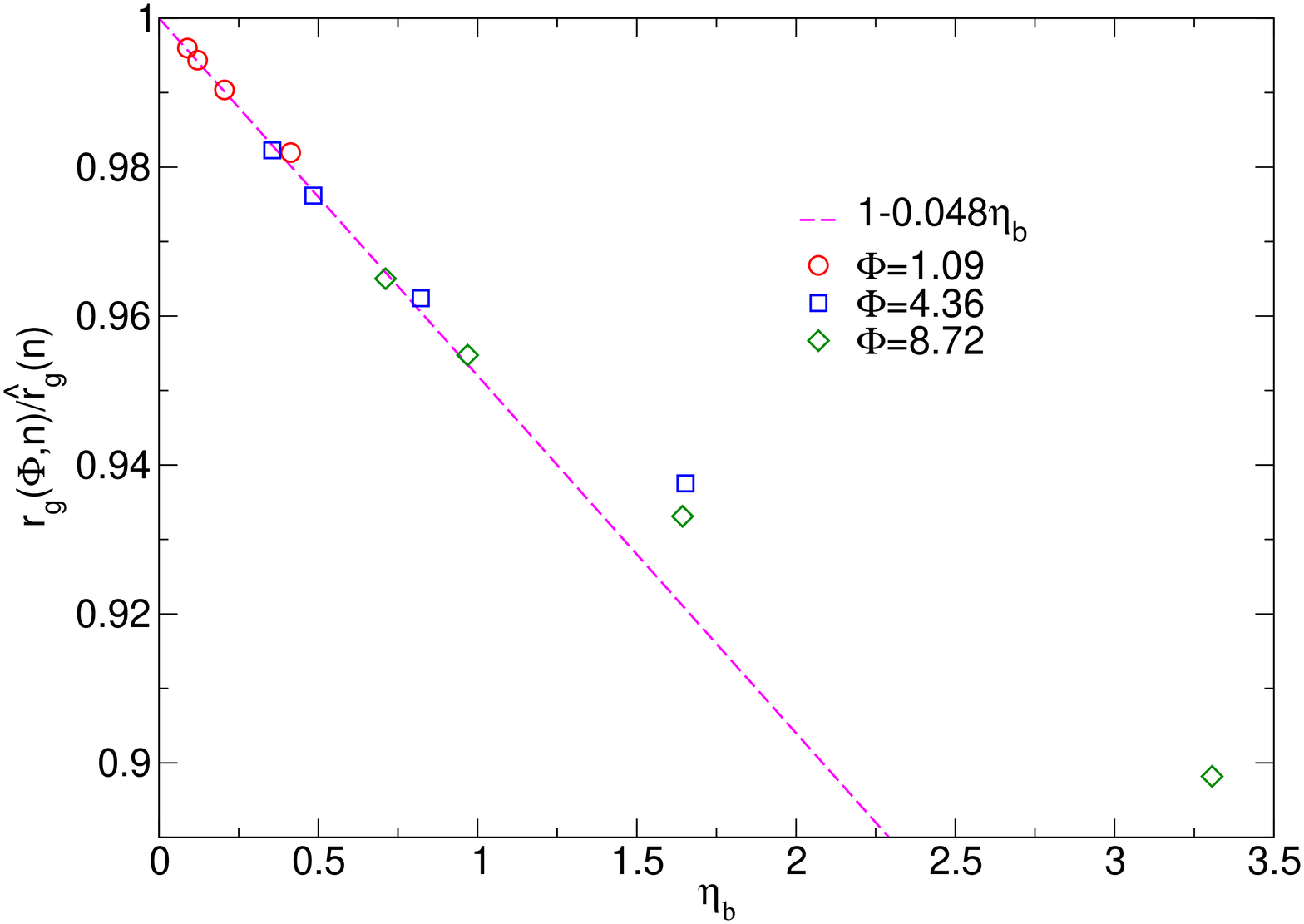,angle=0,width=7truecm} \hspace{0.5truecm} \\
\end{tabular}
\end{center}
\caption{Universal ratio $r_g(\Phi,n)/\hat{r}_g(n)$ as a function
of the blob volume fraction $\eta_b$. The dashed line is a linear fit of the 
data with $\eta_b < 1$. 
}
\label{fig:ratiorg}
\end{figure}

It is also interesting to consider the ratio $r_g(\Phi,n)/\hat{r}_g(n)$,
where $r_g$ is the average blob size in the asymptotic limit $L\to \infty$
(we perform the same extrapolation as done before for the ratios $Q_i$). 
Results for several values of $n$ and $\Phi$ are shown 
in Table~\ref{ratiorg} and plotted in figure~\ref{fig:ratiorg} 
versus the blob volume fraction 
$\eta_b = c_b/c_b^* = 4 \pi \hat{r}_g^3 c_b/3$. 
At least for $\eta_b \lesssim 1$ the data appear to depend only on
$\eta_b$ and to converge to 1 linearly as $\eta_b\to 0$:
$r_g(\Phi,n)/\hat{r}_g(n) \approx 1 - 0.048 \eta_b$. Since $\eta_b\to 0$ 
for $n\to \infty$ at fixed $\Phi$, this result allows us to predict the ratio
$Q(n)$, the average of the $Q_i(n)$ defined above, as $n\to \infty$. 
Indeed, we have 
\begin{equation}
{r_g(\Phi,n) \over R_g(\Phi)} = 
   {r_g(\Phi,n)\over \hat{r}_g(n)}\, {\hat{r}_g(n)\over \hat{R}_g} \,
   {\hat{R}_g \over R_g(\Phi)} \approx {1.03 n^{-\nu}} 
   {\hat{R}_g \over R_g(\Phi)}.
\label{rgoverRg-Phi}
\end{equation}
The ratio $R_g(\Phi)/\hat{R}_g$ has been computed in several works.
\cite{CMP-06-raggi,Pelissetto:2008p1683} For large $\Phi$, 
${\hat{R}_g/ R_g(\Phi)}$ scales \cite{Pelissetto:2008p1683} 
as $0.90\Phi^{0.115}$ so that 
${r_g(\Phi,n)/R_g(\Phi)} \approx 0.93 n^{-\nu} \Phi^{0.115}$. 
Note that scaling (\ref{rgoverRg-Phi}) sets in for quite large
values of $n$ if $\Phi$ is large. For instance, for $\Phi = 8.72$ 
it predicts $n^\nu Q = n^\nu {r_g(\Phi,n)/R_g(\Phi)} \approx 1.23$ for $n\to \infty$, 
since \cite{CMP-06-raggi} 
$R_g(\Phi)/\hat{R}_g \approx 0.84$ for this value of $\Phi$.
Hence, even for $n = 30$, see figure \ref{Q-ratio}, 
we are still far from the asymptotic limit.

\section*{References}

\providecommand{\newblock}{}

\end{document}